\documentclass[12pt]{iopart}
\pdfoutput=1 % if your are submitting a pdflatex (i.e. if you have
\usepackage{graphicx}
\usepackage[T1]{fontenc} % if needed
\usepackage[utf8]{inputenc} %viva à acentuação!
\usepackage{iopams}
\usepackage{bm}
\usepackage{times}
\usepackage{slashed}
\usepackage{epsf}

\begin{document}

\def\micro{\texttt{micrOMEGAs}}

\title[Majorana Dark Matter in Minimal Higgs Portal Models after LUX]{Majorana Dark Matter in Minimal Higgs Portal Models after LUX}

\author{M. Dutra, C. A. de S. Pires and P. S. Rodrigues da Silva}
\address{Departamento de Física, Universidade Federal da
Paraíba, \\ 
Caixa Postal 5008, 58051-970, João Pessoa - PB, Brasil.}

\ead{mdutra@fisica.ufpb.br, cpires@fisica.ufpb.br and psilva@fisica.ufpb.br}

\begin{abstract}
We consider the Singlet Majorana fermion dark matter model, in which the standard model particles interact with the dark sector through the mixing of a singlet scalar and the Higgs boson. In this model both the dark matter and the singlet scalar carry lepton number, the latter being a bilepton. We suppose the existence of a $Z_2$ symmetry, remnant of some high energy local symmetry breaking, that stabilizes the Majorana fermion. We analyzed the parameter space of this model and found that the lepton number symmetry breaking scale, drove by the singlet scalar, is constrained to be within hundreds to thousands of GeV, so as to conform with the observed dark matter relic density.
Finally, we address the direct detection and invisible Higgs decay complementarity, confronting our model with recent LUX and LHC constraints, as well as XENON1T prospects.
\end{abstract}

\section{Introduction}
\label{intro}

Since almost a century ago, a huge ammount of evidence has been accumulated through astrophysics observations~\cite{zwicky1933,rubin1970,machoend2014,bullet2006,mondnaonosrepresenta2007,Dodelson2011}, indicating that the majority of matter in the Universe is of unknown origin.
Such observations, however, refer specifically to the gravitational response of ordinary (or baryonic) matter to this so called Dark Matter (DM), scarcely showing other properties it should comply with, except that it has to be effectively electrically neutral~\footnote{Millicharged DM candidates exist in the literature which are perfectly acceptable though~\cite{Goldberg1986,DeRujula1990,Chuzhoy2008}}, non-baryonic, cosmologically stable and cold at recombination. These are important clues in order to start building a model to describe the DM. Particle Physics is strongly appealing in offering a DM candidate, since many motivating extensions of the standard model of electroweak interactions (SM) possess  new fields and symmetries, which could suitably provide one (or more) stable particle(s) in their spectrum.

The first observational requirement that a DM model has to fulfill is a a set of interactions that can account for the relic density inferred from the power spectrum of the cosmic microwave background radiation (CMB). As it says nothing about the particle nature of the DM candidate, such as its mass, spin or charges, we have to count on the possibility of detecting a signal from a DM particle and gathering more information in order to put this particle under siege. Such a signal should be identified in underground detectors (direct detection search), in sattelites or ground-based telescopes (indirect detection search) as well as in colliders or accelerators (collider search). There are many experiments searching for DM candidates, especially the weakly interacting massive particles (WIMPs). Currently, no unambiguous positive DM signal was announced by the experiments, but limits were put on its scattering cross-section off nucleon and production rate. 

Any viable particle physics model that addresses the DM issue must comply with these bounds and may be constructed in an top-down or a bottom-up way. The top-down models intend to explain open particle physics problems in a fundamental way while having a DM candidate, such as supersymmetric and extra-dimension models. The bottom-up models intend to minimally explain the DM physics, with just a few free parameters and strong predictions, and are more easily ruled out by the experiments. This strategy, that we will adopt here, allows us to identify what initial hypothesis can be modified in order to agree with the experimental limits.

Models in which the interaction between DM and standard particles is realized through scalar exchanges, the Higgs portal, are phenomenologically interesting since the discovery of the Higgs-like scalar at the LHC~\cite{higgs1,higgs2}. As its couplings and branching ratios are being measured, the Higgs physics can constrain the parameter space of such DM models. Here, we focus on the fermionic DM in a minimal Higgs portal model. Fermionic DM particles may be Majorana or Dirac particles, differing in their relic abundance which is twice larger for Dirac fermions, to account for the fact there are two distinct particles annihilating~\cite{Fedderke:2014wda}. Also, the scattering cross-section off nucleon of a Majorana fermion is twice larger relative to a Dirac fermion~\cite{Baek:2014jga}. In many respects there is no significant difference between Dirac and Majorana fermion DM candidates though, as pointed out in previous works~\cite{Baek:2011aa,Fedderke:2014wda,LopezHonorez:2012kv,Okada:2013rha,Boeckel:2007tz}.
 
Singlet Dirac DM candidates were vastly explored in the context of Higgs portal models after the Higgs discovery~\cite{Ghorbani:2014qpa,Bagherian:2014iia,Franarin:2014yua,Fairbairn:2013uta,Chua:2013zpa,Baek:2011aa}. It is known that a lighter than 100~GeV Dirac DM is excluded by relic, direct detection and collider constraints, except at the resonance region~\cite{Kim:2008pp,Baek:2012uj}. Although less considered, a Majorana DM in renormalizable Higgs portal models were also pointed out as viable DM candidates~\cite{Esch:2013rta,Okada:2013rha,LopezHonorez:2012kv}. 

The effective field theory (EFT) approach of the Majorana fermion DM was considered in~\cite{Matsumoto:2014rxa,Fedderke:2014wda,LopezHonorez:2012kv}. Within EFT approach, the fermionic DM Higgs portal CP-conserving scenario is excluded by direct detection search for the 60~GeV - 2~TeV DM mass range. Nevertheless, it remains viable if we consider UV completions such as a resonant or an indirect Higgs portal~\cite{LopezHonorez:2012kv}.
Since we do not know the mass scales of the possible intermediate particles, we can not know if an EFT approach is valid in the context of Higgs Portal DM models. Indeed, even though renormalizable models must agree with EFT models in the limit of heavy mediators, some degrees of freedom concerning the mediator phenomenology are only appreciable in concrete models. The limitations of the EFT approach to Higgs Portal DM models, concerning the LHC and LUX constraints ~\cite{lux}, were considered in refs.~\cite{Baek:2014jga,Busoni:2013lha}. Also, the implications of the 125GeV scalar for Higgs Portal DM models were studied in a model independent way in ref.~\cite{Djouadi:2011aa}, considering generic scalar, Majorana fermion and vector DM candidates. They found that a non-renormalizable Majorana fermion DM is excluded by XENON100, if the Higgs-DM coupling strength is larger than $10^{-3}$.

In this work, we consider a sterile Majorana neutral fermion as DM candidate in a renormalizable CP-conserving Higgs portal model~\cite{deS.Pires:2010fu}, under the light of the recent experimental limits. We suppose the existence of a discrete $Z_2$ symmetry responsible for the Majorana fermion stability, that is expected as a remnant of some spontaneously broken gauge symmetry at a high scale not relevant for the low energy physics. The mediator between the Majorana fermion and the standard particles is a complex singlet scalar that develops a nonzero vacuum expectation value (vev), thus breaking lepton number symmetry. 

This work is the first consideration of the LUX impact on the parameter space of a concrete Majorana DM Higgs portal model. We also consider the XENON1T prospect bound ~\cite{2012arXiv1206.6288A}, the invisible Higgs decay constraint and comment how the extension of the scalar sector alters the Higgs self-couplings, showing that its future measurement will put important bounds on our parameter space.

This work is organized as follows: in the next section we will present our model. In section~\ref{s3} we compute the relic density and scattering cross section off nucleon for the Majorana fermion, in order to constrain our parameter space and discuss how the free parameters influence this observables. We also comment the impact of the future measurements of the Higgs self-couplings on the free parameters and the possibility of having a dark radiation candidate in our scenario. Our main results are presented in section~\ref{s4}, where we update the parameter space taking into account the invisible Higgs decay, the LUX and the XENON1T bounds, considering separately the scalar extension and the mass spectrum parts. Finally, in section~\ref{s5} we give our conclusions. 

\section{The model} \label{s2}

The model we consider here consists of a singlet Majorana right-handed neutral fermion, $N_R$, as a DM candidate, whose interaction with the SM particles is through the Higgs portal by including a singlet scalar in the model, $\sigma$, which mix to the SM Higgs doublet, $\phi$. This Majorana fermion is sterile and stable, a requisite for a DM candidate, when non-trivially transforming under a $Z_2$ symmetry, which forbids it to mix to the light neutrinos and makes it interact only with the singlet scalar field.  

We built our Lagrangian assuming lepton number conservation. Considering that $N_R$ carries one unit of lepton number, no explicit mass term for it is present in the lagrangian. Its mass is generated from an Yukawa term involving the $Z_2$-even singlet bilepton, $\sigma$, when it acquires a nonzero vev and break the lepton number symmetry. This assumptions reduce the number of free parameters in our model, in comparison with the Majorana fermion models we referenced. 

The most general renormalizable Lagrangian to be added to the SM, under the above assumptions, is given by
\begin{equation}  
\mathcal{L} \supset \mathcal{L}_{kin}(N_R ,\sigma) - \lambda_{N}(\overline{N^c_R}N_R \sigma + \overline{N_R}N^c_R \sigma^*)
- V(\phi,\sigma) ,  
\label{lag}
\end{equation}
where $\mathcal{L}_{kin}(N_R ,\sigma)$ is the kinetic term for the new singlet fields and the scalar potential is
\begin{eqnarray} \label{potencial}
V(\phi,\sigma) = && \mu_\phi^2 \phi^\dagger \phi + \lambda_\phi (\phi^\dagger \phi)^2 + 
		  \mu_\sigma^2 \sigma^* \sigma + \lambda_\sigma (\sigma^* \sigma)^2 \nonumber \\
		  && + \lambda_{\phi \sigma}(\phi^\dagger \phi)(\sigma^* \sigma)  .
\end{eqnarray}

After spontaneous breaking of electroweak and lepton number symmetries, when $\langle \phi^0\rangle\equiv v_\phi/\sqrt{2}$ and $\langle \sigma\rangle\equiv v_\sigma/\sqrt{2}$, the above potential leads to the minimum conditions:
\begin{equation}
v_\phi^2= - \frac{\mu_\phi^2 + \lambda_{\phi \sigma} \langle 
\sigma^2 \rangle_0}{\lambda_\phi}
\label{vi1}
\end{equation}
and
\begin{equation}
v_\sigma^2  = - \frac{\mu_\sigma^2 + \lambda_{\phi \sigma} 
\langle \phi^2 \rangle_0}{\lambda_\sigma}\,.
\label{vi2}
\end{equation}
The physical spectrum is then obtained by shifting the scalar fields to the physical fields, that can be written in the unitary gauge as,
\begin{equation} 
\phi = \frac{1}{\sqrt{2}} \left( \begin{array}{c} 0 \\ v_\phi + R_\phi 
\end{array} \right)  
\end{equation}	
and
\begin{equation}
\sigma = \frac{v_\sigma + R_\sigma + i J}{\sqrt{2}}\,.
\end{equation}

Observe that the breaking of lepton number symmetry implies the existence of a Majoron in the spectrum, $J$, which was shown to be a phenomenologically safe Goldstone boson~\cite{deS.Pires:2010fu}. The mass matrix of the massive scalar fields is given by
\begin{equation} \label{mass}
M^2 = \left( \begin{array}{cc}
2 \lambda_\phi v_\phi^2 & \lambda_{\phi \sigma} v_\phi v_\sigma  \\
\lambda_{\phi \sigma} v_\phi v_\sigma & 2 \lambda_\sigma v_\sigma^2  \\    
\end{array}\right).\quad 
\end{equation}	

After diagonalization, we have the physical states
\begin{eqnarray}
 H = c_\alpha R_\phi + s_\alpha R_\sigma \nonumber \\
 S = - s_\alpha R_\phi + c_\alpha R_\sigma,
  \label{mix}
\end{eqnarray}
where $c_\alpha$ and $s_\alpha$ are the cosine and sine of the mixing angle $\alpha$, determined by
\begin{equation} \label{tan}
 \tan{2\alpha} = \frac{\lambda_{\phi \sigma} v_\phi v_\sigma}{\lambda_\phi v_\phi^2 - \lambda_\sigma v_\sigma^2}.
\end{equation}

Since the discovery of a scalar resonance in the CMS and ATLAS experiments at LHC~\cite{higgs1,higgs2}, with mass $M_H=125$~GeV, and whose interactions indicate it is most probably the SM Higgs boson, we can be sure that this scalar cannot be a singlet under $SU(2)_L$. In our case, it has to come from the doublet-like scalar, $R_\phi$, implying that the mixing angle in eq.~(\ref{mix}) has to be small. This is further corroborated by the null search results for a lighter than 125~GeV scalar, once the coupling of SM fields and the scalar singlet are suppressed by $s_\alpha$. We then identify the mass eigenstates by
\begin{equation} \label{mh}
M_H^2 = 2( \lambda_\phi v_\phi^2 c_\alpha^2 + \lambda_\sigma v_\sigma^2 s_\alpha^2 + \lambda_{\phi \sigma} v_\phi v_\sigma s_\alpha^2 c_\alpha^2 )
\end{equation}
for the Higgs scalar and
\begin{equation} \label{ms}
M_S^2 = 2( \lambda_\sigma v_\sigma^2 c_\alpha^2 + \lambda_\phi v_\phi^2 s_\alpha^2 - \lambda_{\phi \sigma} v_\phi v_\sigma s_\alpha^2 c_\alpha^2 )\,.
\end{equation}
for the singlet scalar.

In terms of the mass eigenvalues, the mixing angle is defined by
\begin{equation}
 \sin{2\alpha} = \frac{2 \lambda_{\phi \sigma} v_\phi v_\sigma}{M_H^2 - M_S^2},
\end{equation}
and the mixing coupling is given by
\begin{equation}
 \lambda_{\phi \sigma} = s_\alpha c_\alpha \frac{M_H^2-M_S^2}{v_\phi v_\sigma}.
\end{equation}
As a consequence of the mixing (eq.~\ref{mix}), all the standard couplings to the Higgs boson will be rescaled in our model by $c_\alpha$. It allows us to probe this model at colliders, looking for deviations from the standard interactions with the Higgs boson. 
For further numerical convenience, we can solve the eqs.~(\ref{mh}) and (\ref{ms}) for the quartic couplings, giving
\begin{equation}\label{lfls}
 \lambda_\phi = \frac{c_\alpha^2 M_H^2 + s_\alpha^2 M_S^2 }{2 v_\phi^2}   \hspace{2cm}  \lambda_\sigma =  \frac{c_\alpha^2 M_S^2 + s_\alpha^2 M_H^2 }{2 v_\sigma^2}.
\end{equation}

The tree level stability of the scalar potential is guaranteed once $\lambda_\phi > 0$, $\lambda_\sigma > 0$ and $-2\sqrt{\lambda_\phi \lambda_\sigma} < \lambda_{\phi \sigma} < 2\sqrt{\lambda_\phi \lambda_\sigma}$, which we will use throughout this work~\footnote{The stability of a somewhat similar model at one loop level, where the scalar is real and the DM is a Dirac fermion, was considered in Ref.~\cite{Baek:2012uj} while, in Ref.~\cite{Fairbairn:2013uta}, the DM issue was also analyzed along with the possibility of a first order phase transition. This last feature not reproducible by our model.}

\begin{figure}[ht]
 \centering
 \includegraphics[scale=0.5]{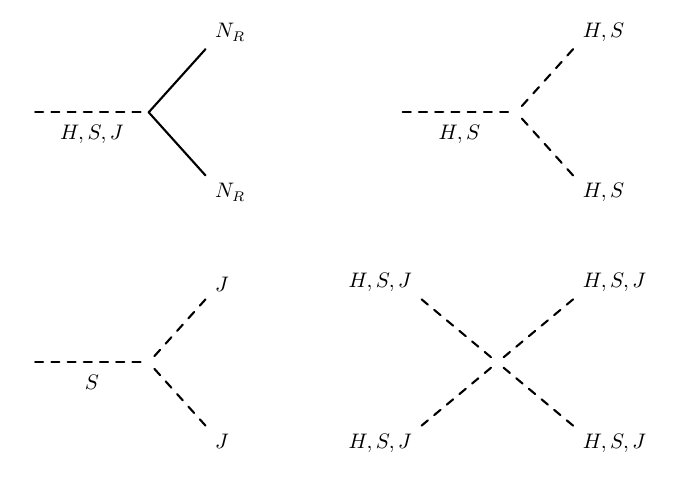}
 \caption{New interactions of the SM extension we are considering.}
 \label{vertices1}
\end{figure}

To complete the spectrum we write down the Majorana fermion mass after the spontaneous breaking of lepton number,
\begin{equation}\label{mn}
  M_N = \sqrt{2} \lambda_N v_\sigma.
\end{equation}
The non-standard interactions introduced by this extension of SM are presented in figure~\ref{vertices1}.

In summary, from the twelve free parameters, $\lambda_N, \mu_\phi^2, \mu_\sigma^2, \lambda_\phi, \lambda_\sigma, \lambda_{\phi \sigma}, M_N, M_S, M_H, v_\phi, v_\sigma$ and $\alpha$, when we consider the six constraints, eqs.~(\ref{vi1},~\ref{vi2},~\ref{tan},~\ref{mh},~\ref{ms} and~\ref{mn}), along with the measured Higgs mass $M_H = 125$~GeV and the electroweak scale, $v_\phi = 246.22$~GeV, we may choose to keep the remaining independent set of parameters as, $M_N, M_S, v_\sigma$ and $\alpha$.
In what follows, we discuss the role of each of these four free parameters in the observables.

\section{Constraining the parameter space}\label{s3}

Our parameter space is constrained by many requirements. First of all, the Majorana fermion must be sufficiently abundant in order to be a viable DM candidate. Here, we use the latest results of the Planck telescope~\cite{Ade:2013zuv}. We computed the relic density and the scattering cross section off nucleon (Direct Detection) for the Majorana fermion, using the numerical package \micro~\cite{micro} and compared it with the LUX bounds~\cite{lux}, the strongest we have to date, as well as with the XENON1T prospects~\cite{2012arXiv1206.6288A}. In this section, we will study how the free parameters affect these observables.

In order to carry out numerical scans on the parameter space, we randomly sampled all the free parameters in the ranges shown in table~\ref{sample}. In ref.~\cite{Walker:2013hka}, we find that the unitarity and perturbativity bounds together require $M_N, M_S < 3$~TeV and $v_\sigma < 2.4$~TeV \footnote{Although these results are based on the assumption that the Majorana mass terms for fermionic DM are absent at tree level, we can naively assume they may be used in our case.}.

Throughout this work, all the points in the scans will satisfy these conditions:
\begin{itemize}
 \item $\alpha$ goes from -0.5735 to 0.5735, to be consistent with the LHC bound for the mixing ($|cos(\alpha)|>0.84$~\cite{Profumo:2014opa});
 \item $\lambda_\phi > 0$, $\lambda_\sigma > 0$ (automatic, see eq.~\ref{lfls}) and $-2\sqrt{\lambda_\phi \lambda_\sigma} < \lambda_{\phi \sigma} < 2\sqrt{\lambda_\phi \lambda_\sigma}$ for a potential bounded from below;
 \item Perturbativity: $\lambda_{\phi \sigma}, \lambda_\phi, \lambda_\sigma < 4\pi$.
 \end{itemize}
and keep $m_H =$~125GeV and $v_\sigma =$~246GeV.

\begin{table} 
 \centering
\scalebox{0.8}
{
 \begin{tabular}{|c|c|c|c|}
\hline
        $M_N$ (GeV)       &    $M_S$ (GeV)     & $v_\sigma$ (GeV)         & $\alpha$     \\
\hline
\hline   [1,2000]          &    [0.1,2000]       & [100,2000]                & [-0.5735,0.5735]     \\  
\hline
\end{tabular}
}
\caption{Ranges for the free parameters used in our numerical scans.}
\label{sample}
\end{table}

\begin{figure}[ht]
\begin{minipage}[t]{0.45\textwidth}
\includegraphics[width=\textwidth]{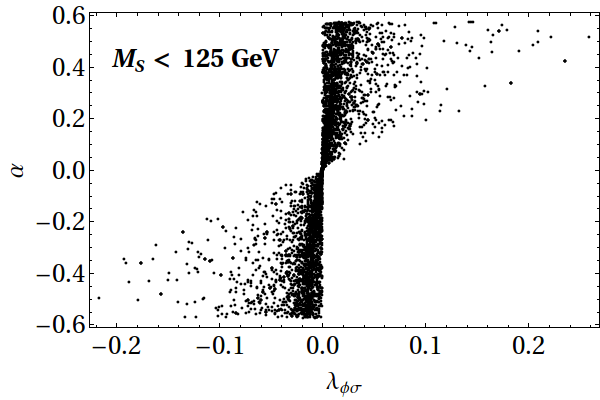}
\end{minipage}\hfill
\begin{minipage}[t]{0.45\textwidth}
\includegraphics[width=\textwidth]{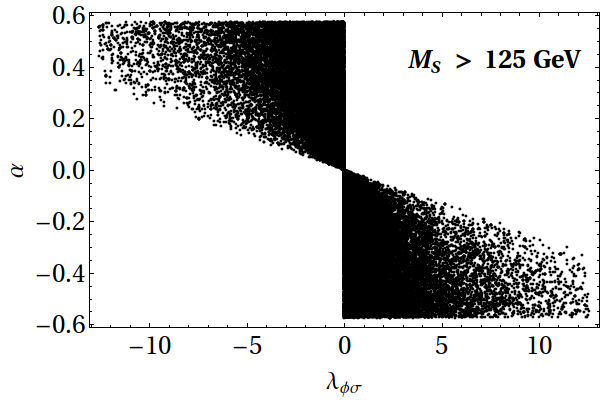}
\end{minipage}
\caption{Projections of the parameter space onto the ($\lambda_{\phi \sigma}$,$\alpha$) plane, for a lighter (left) and heavier (right) than Higgs singlet scalar. Note that the former case is possible only for a very weak mixing coupling, there is no point for $|\lambda_{\phi \sigma}|$ much greater than 0.2 under the conditions we are considering and the chose range of free parameters.}
\label{lfsMsig}
\end{figure}

It is instructive to emphasize that this parametrization allows us to consider the two possible scalar mass hierarchies: for a positive (negative) $\lambda_{\phi \sigma}$, $\alpha > 0$ ($\alpha < 0$) ensures that $M_S < M_H$ and $\alpha < 0$ ($\alpha > 0$) ensures that $M_S > M_H$, as we can see in figure~\ref{lfsMsig}. In this figure, we see the projection of all the parameter space (with the free parameters varying according to the Table~\ref{sample}) onto the ($\lambda_{\phi \sigma}$,$\alpha$) plane.

An important feature of this Higgs Portal model is that the ongoing characterization of the discovered Higgs boson can provide valuable constraints on the free parameters. One example of this is the future measurement of the Higgs self-couplings~\cite{Barr:2014sga,Conway:2014gaa,Tian:2013yda,Goertz:2013eka}, which we intend to examine next.
The 3-Higgs and 4-Higgs self-couplings in our model are given by

\begin{equation}
 \lambda_{HHH} = \frac{M_H^2}{2 v_\phi}\left(c_\alpha^3 - s_\alpha^3 \frac{v_\phi}{v_\sigma}\right) 
\end{equation}
and
\begin{equation}
 \lambda_{HHHH} = c_\alpha^4 \frac{\lambda_\phi}{4} + s_\alpha^4 \frac{\lambda_\sigma}{4} + s_\alpha^2 c_\alpha^2 \frac{\lambda_{\phi \sigma}}{4}.
\end{equation}

\begin{figure}[h]
\centering
\begin{minipage}[t]{0.45\textwidth}
\includegraphics[width=\textwidth]{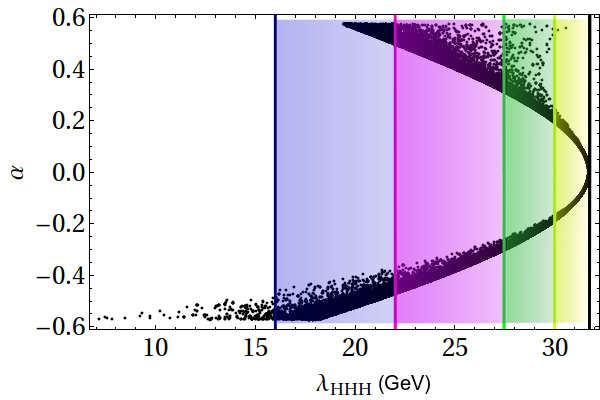}
\end{minipage}\hfill
\begin{minipage}[t]{0.45\textwidth}
\includegraphics[width=\textwidth]{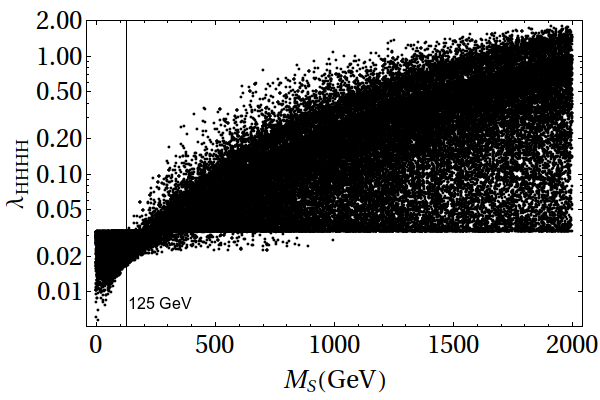}
\end{minipage}
\caption{3-Higgs (left) and 4-Higgs (right) self-couplings as functions of the mixing angle and singlet scalar mass, respectively. Note that deviations from the standard values give us information about the intensity of the scalar mixing as well as the scalar mass hierarchy, if $M_S$ is smaller or higher than 125~GeV as we emphasize at the right panel. In the left panel, the coulored bands are prospect deviations of $\pm50\%$ (TLEP500 / HL-LHC-min), in blue, $\pm30\%$ (TLEP240 / CEPC HL-LHC-max), in magenta, $\pm13\%$ (ILC), in green, and $\pm5\%$ (hadron100TeV), in yellow, from the SM predicted value (black line).}
\label{selfH}
\end{figure}

In figure~\ref{selfH}, we see the projections of the parameter space onto the ($\lambda_{HHH},\alpha$) (left) and ($M_S,\lambda_{HHHH}$) (right) planes. For the 3-H self coupling, we see that deviations from the SM value ($\lambda_{HHH} = 31.73$~GeV) give us information about the intensity of the scalar mixing. Following~\cite{Profumo:2014opa}, we see in the 3-Higgs self-coupling projection the bands corresponding to the future collider sensitivities: deviations from the SM predicted value of $\pm50\%$ (TLEP500 / HL-LHC-min) are in blue, $\pm30\%$ (TLEP240 / CEPC HL-LHC-max) in magenta, $\pm13\%$ (ILC) in green and $\pm5\%$ (hadron100TeV) in yellow. For the 4-Higgs self coupling, we see that significant deviations from the SM value ($\lambda_{HHHH} = 0.032$) will be able to establish the scalar mass hierarchy in our scenario. With this picture in mind we can follow with the analysis of the DM constraints on the parameter space.

\subsection{Relic Density}

\begin{figure}[t]
\centering
\begin{minipage}{0.25\textwidth}
\includegraphics[scale=0.2]{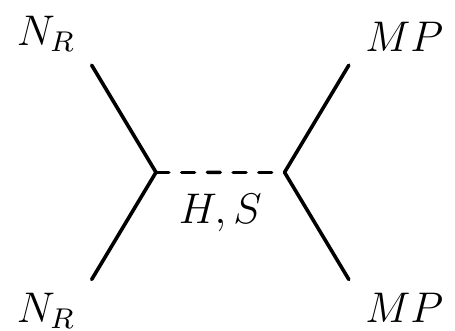}
\end{minipage}\hfill
\begin{minipage}{0.25\textwidth}
\includegraphics[scale=0.2]{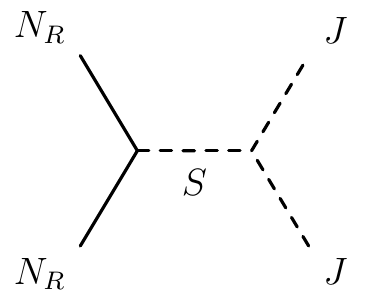}
\end{minipage}\hfill 
\begin{minipage}{0.25\textwidth}
\includegraphics[scale=0.2]{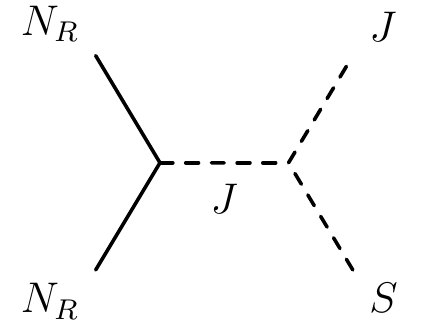}
\end{minipage}\hfill
\begin{minipage}{0.25\textwidth}
\includegraphics[scale=0.2]{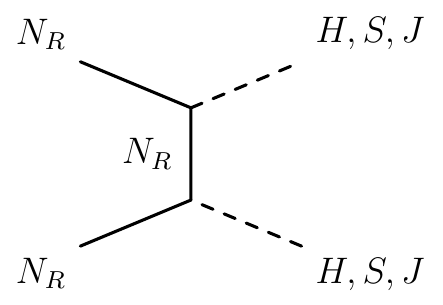}
\end{minipage} 
\caption{Annihilation channels relevant for the Majorana fermion relic density.}
\label{aniq1}
\end{figure}

In this work we are assuming dark matter thermal production, thus the Majorana fermion relic density is determined by calculating the thermal averaged annihilation cross section for the processes shown in figure~\ref{aniq1}. 

As the dark matter in our scenario interacts just by scalar exchanges, a characteristic behavior of the relic density is one sharply peaked region at $M_N \approx M_H/2 \approx 63$~GeV, corresponding to the Higgs production and depending on the scalar mixture, and another one at $M_N \approx M_S/2$, corresponding to the singlet scalar production. As an asymptotic behavior, we expect a gradual decrease in the relic density, as more channels turn out kinetically allowed for more massive Majorana fermions. 

\begin{figure}[t]
 \centering
 \includegraphics[scale=0.28]{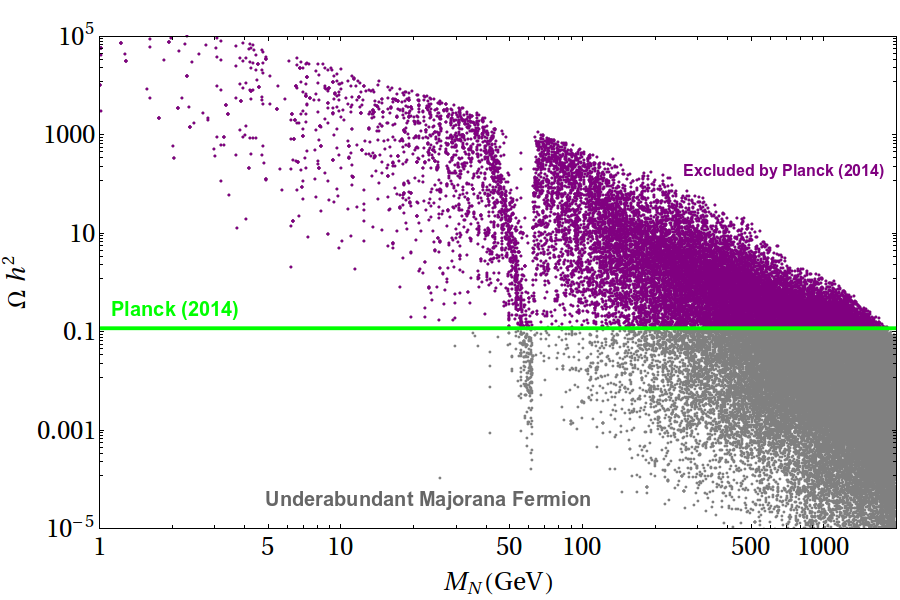}
 \caption{Relic density as a function of the Majorana fermion mass. The green interval represents the Planck bound on the relic density, for any DM mass. The purple points overclose the Universe and will not be considered in our analysis. The gray points indicate regions of our parameter space in wich the Majorana fermion is not sufficient to account for the total DM content of the Universe.}
 \label{omega}
\end{figure}

In figure~\ref{omega}, we show the projection of the parameter space onto the ($M_N,\Omega h^2$) plane. Our parameter scan shows that there is no set of free parameters allowing for sufficiently abundant Majorana DM lighter than few tens of GeV, in agreement with~\cite{Matsumoto:2014rxa}. Considering that the lepton number symmetry was spontaneously broken at some scale between 100-2000~GeV, the singlet scalar mass from 0.1-2000~GeV and a mixing angle ensuring $|\cos(\alpha)|>0.84$, we can have a Majorana fermion as a DM candidate for masses from 30~GeV to 2~TeV.

\begin{figure}[h]
\centering
\minipage{0.4\textwidth}
  \includegraphics[width=\linewidth]{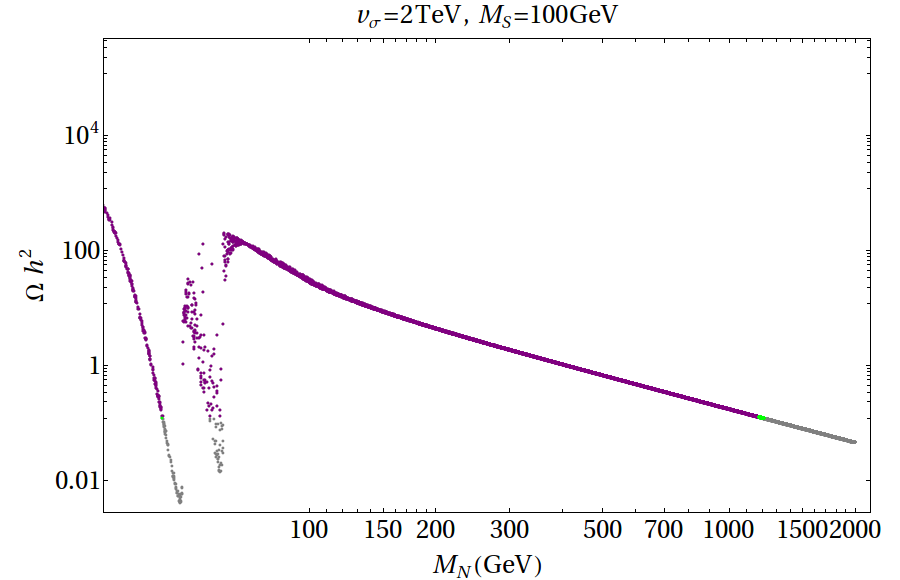}
\endminipage
\minipage{0.4\textwidth}
  \includegraphics[width=\linewidth]{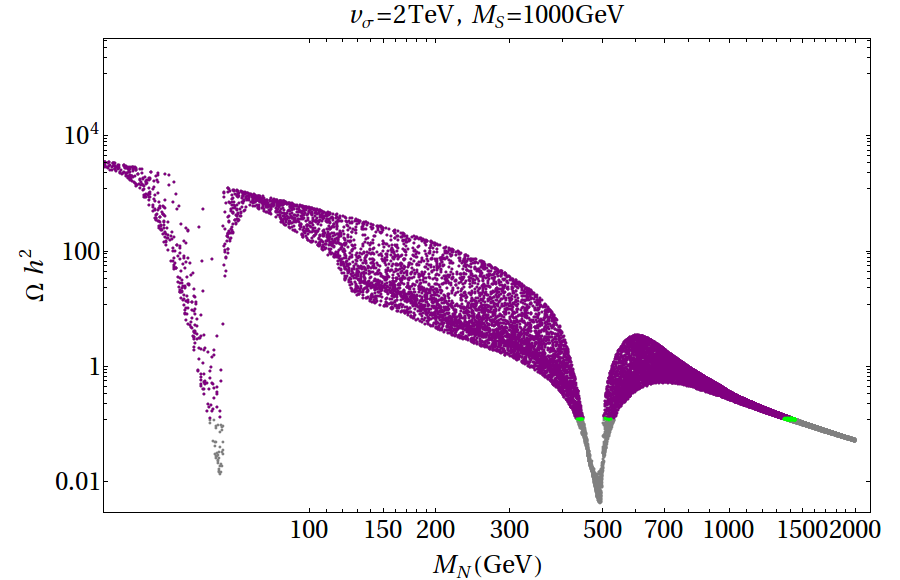}
\endminipage
\caption{Singlet scalar resonances, for all values of the mixing angle. Notice that the Planck-allowed points (green) are always near the scalar resonances and where the abundance curve falls down. It is a characteristic behavior of this model, as we will see in more detail in what follows. Same color code of figure~\ref{omega}.}
\label{resonances}
\end{figure}

The resonance due to the Higgs exchange, at $M_N \sim 62.5$~GeV, is clear in these plots, but if it is going to reach the observed relic density depends mainly on the mixing angle value. Stronger the scalar mixing, more intense the annihilation in standard particles and consequently smaller the relic density for a Majorana fermion at such mass scale. In figure~\ref{resonances} we show two particular values of $M_S$ that set different singlet scalar resonances.

\begin{figure}[h]
\centering
\minipage{0.4\textwidth}
  \includegraphics[width=\linewidth]{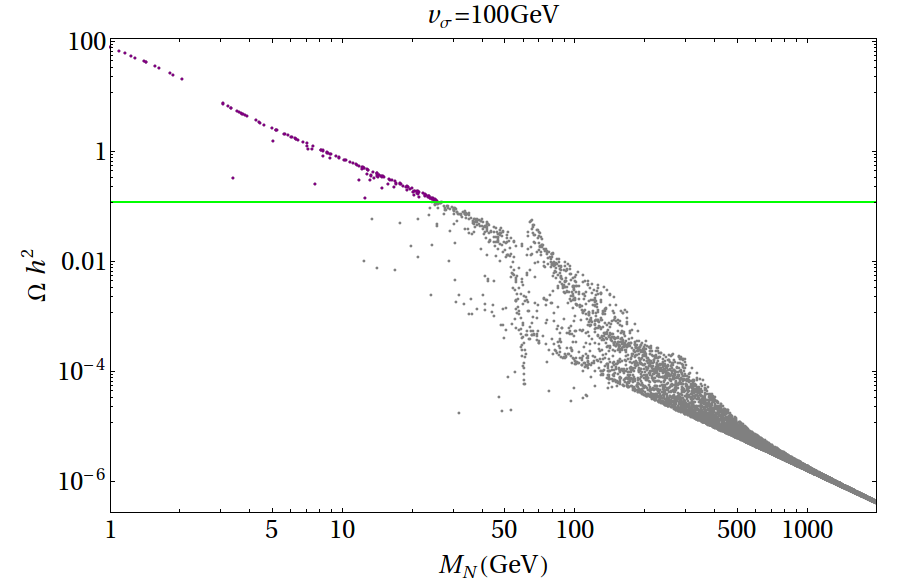}
\endminipage
\minipage{0.4\textwidth}
  \includegraphics[width=\linewidth]{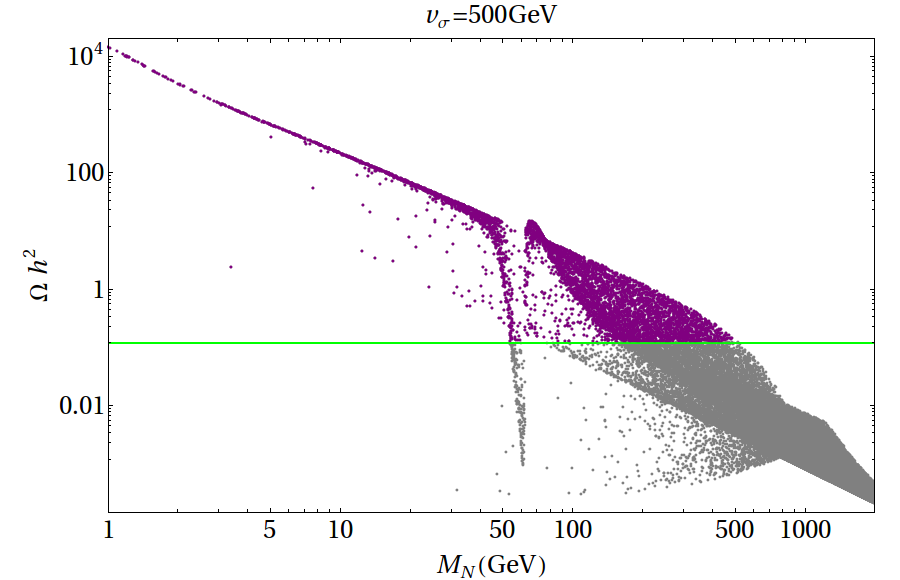}
\endminipage\hspace{1cm} 
\minipage{0.4\textwidth}
  \includegraphics[width=\linewidth]{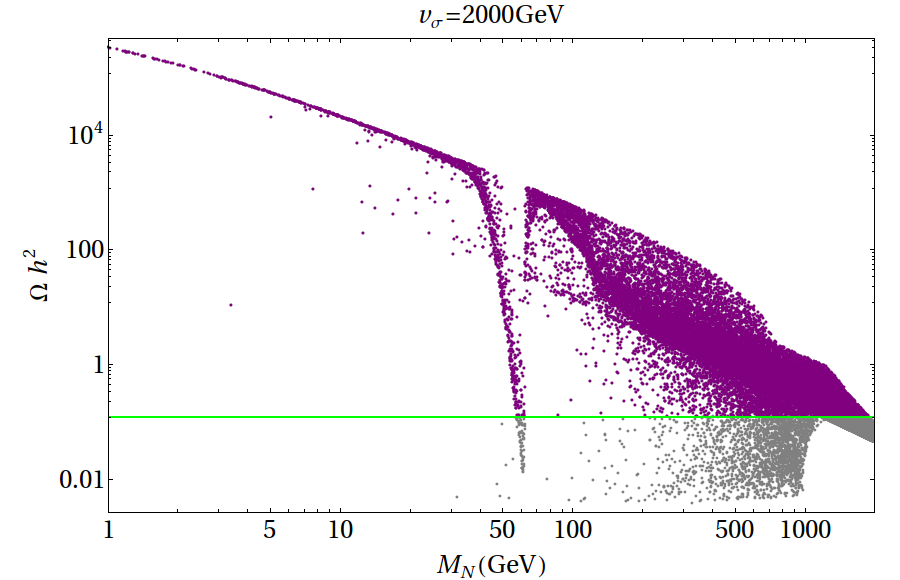}
\endminipage
\minipage{0.4\textwidth}
  \includegraphics[width=\linewidth]{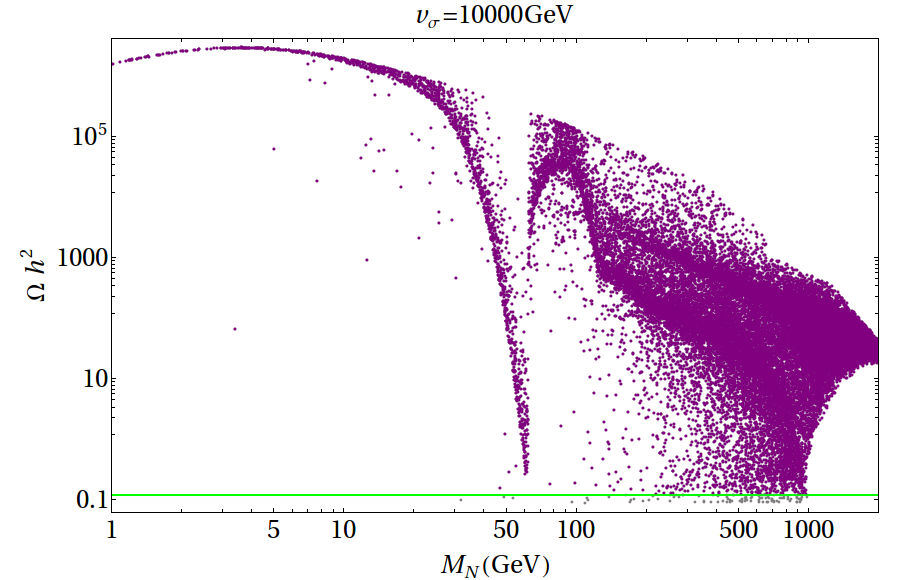}
\endminipage
\caption{Influence of the lepton number symmetry breaking scale on the relic density. Same color code of figure~\ref{omega}.}
\label{vev}
\end{figure}

Let us now see how the lepton number symmetry breaking scale, $v_\sigma$, changes the relic density. In figure~\ref{vev}, we see that there is a global increase in the relic density according to the increase of $v_\sigma$, due mainly to the interaction of the Majorana fermion with H, S and J, which is inversely proportional to $v_\sigma$. In view of this, we have an upper and a lower allowed scale for the global symmetry breaking to occur in this DM scenario. The plots in figure~\ref{vev} comprehend all the $M_S$ values, then all the singlet scalar resonances. As we are taking singlet scalars up to 2~TeV, the resonances due to their productions go up to 1~TeV. 

We can now look at the parameter space under the light of direct detection experiments, which we do next.

\subsection{Direct Detection}

The interaction between the Majorana fermion and the quarks are realized via t-channel scalar exchanges. So, the scattering cross section is always spin-independent (SI) and is given by
\begin{equation}
 \sigma_{NNqq} = \frac{|\mathcal{M}_{NNqq}|^2}{128 \pi^2 E_{cm}^2} \left(1-\frac{4 M_q^2}{E_{cm}^2}\right)^{1/2} \left(1-\frac{4 M_N^2}{E_{cm}^2}\right)^{-1/2}
\end{equation}

The scattering amplitude is the sum of each scalar contribution:
\begin{equation} \label{amplitude}
 \mathcal{M}_{NNqq} = \mathcal{M}_H + \mathcal{M}_S \varpropto i\left(\frac{M_q M_N}{v_\phi v_\sigma}s_\alpha c_\alpha \right) \left( \frac{1}{M_H^2} -  \frac{1}{M_S^2} \right),
\end{equation} 
where we have taken the low energy limit in which the masses of the virtual particles are much greater than their momenta. This minus sign comes from the diagonalization of the scalar mass eigenstates (eq.~\ref{mix}).

\begin{figure}[ht]
\centering
\includegraphics[scale=0.3]{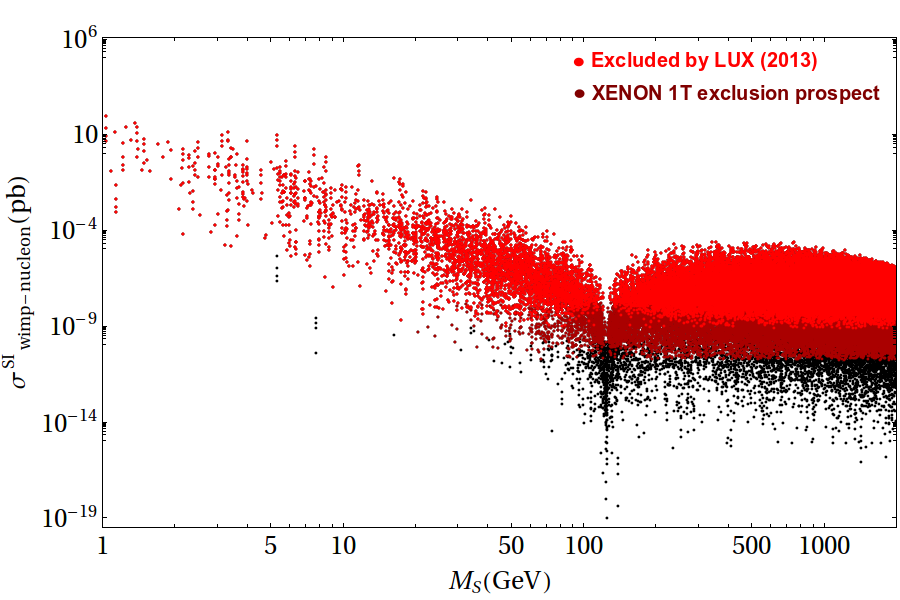}
\caption{Scattering cross section off nucleon as a function of the singlet scalar mass. The red points are already excluded by the LUX results. The dark red points will be excluded by XENON 1T if no signal is confirmed in the next few years. Notice that the case of degenerate scalars can evades the direct detection bounds for all the free parameter values and very light singlet scalars are not favored since they enhance the scattering cross section.}
\label{sigmaMsig}
\end{figure}

We present a scatter plot for the scattering cross section off nucleon in figure~\ref{sigmaMsig}, that shows the well known interference effect in the amplitude near the Higgs mass value, as can be expected from eq.~\ref{amplitude}. This is why the case of $M_S\sim$125~GeV evades the LUX constraint, as well as the future XENON 1T constraint, for all the free parameter values considered. We also see the enhancement of the scattering cross section provided by low scalar masses.

\begin{figure}[!htb]
\minipage{0.35\textwidth}
  \includegraphics[width=\linewidth]{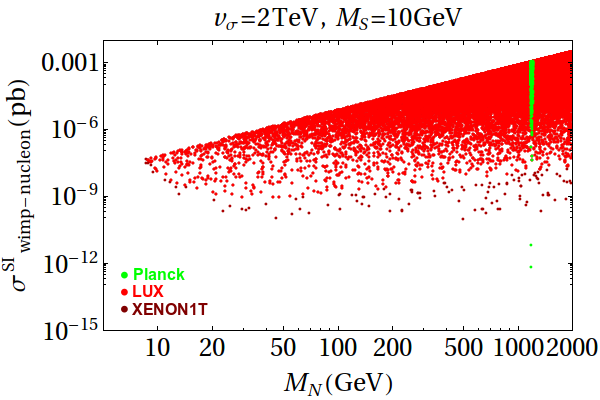}
\endminipage 
\minipage{0.35\textwidth}
  \includegraphics[width=\linewidth]{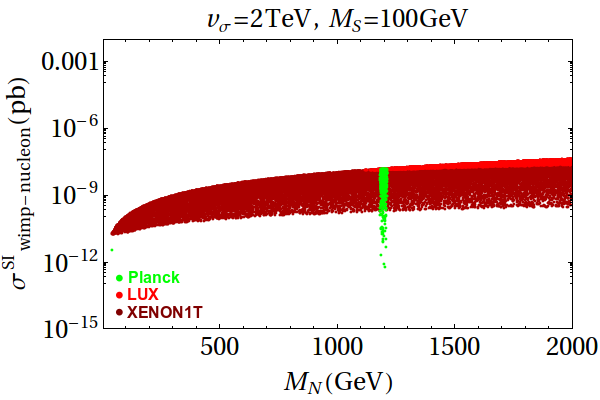}
\endminipage
\minipage{0.35\textwidth}
  \includegraphics[width=\linewidth]{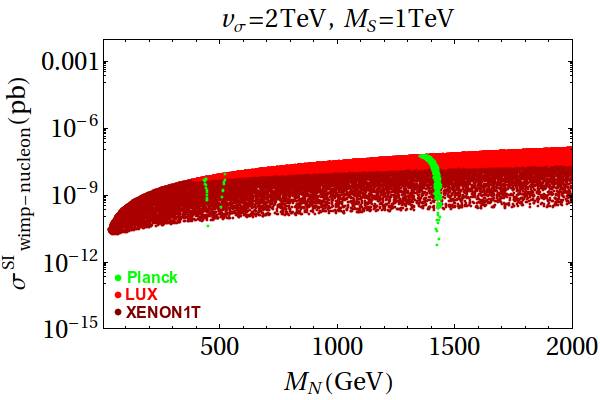}
\endminipage
\caption{Scattering cross section as a function of the Majorana fermion for all values of the mixing angle and certain values of the singlet scalar mass. We see that a lighter than Higgs singlet scalar is quite ruled out by LUX (left). The cases of a quite degenerate (middle) and a TeV scale (right) singlet scalar can evade the direct detection bounds, but will be strongly constrained by the XENON1T results.}
\label{msigfixed}
\end{figure}

In figure~\ref{msigfixed} we show some plots that clarify the effects of the singlet scalar mass on the WIMP-nucleon scattering cross section, taking into account the points that are excluded by the LUX results (red), and the points within the Planck interval (gree) for the relic density, as well as the XENON1T exclusion prospect (dark red). 

Although there exist the possibility of a lighter than Higgs singlet scalar in this model~\cite{Clarke:2013aya}, this case is quite ruled out by LUX in our DM scenario, as we can see in figure~\ref{msigfixed} (left). Henceforth, we will focus on the case of a singlet scalar heavier than the Higgs boson. As we have already remarked, the case in which the scalars are quite degenerate leads to a cancellation effect in the scattering amplitude, evading the LUX constraint and even the XENON1T future constraint (figure~\ref{msigfixed}, middle). Notice that XENON1T will be restrictive even for a TeV scale singlet scalar (figure~\ref{msigfixed}, right).   

The inviability of very light singlet scalars in this model has an interesting consequence. As we have seen, there exists in our spectrum a Majoron, J, that is a massless pseudoscalar. Cosmologically, this particle will naturally contribute to the radiation energy density as long as its interactions make it decouple from the thermal bath. As there are evidences for an unknown contribution for the relativistic degrees of freedom at recombination, the so called dark radiation, we have checked the possibility of J address this question in our dark matter scenario. 

The Majoron can be a radiation at the recombination epoch if it decouples just before the muon annihilation~\cite{Weinberg:2013kea} and in that case, roughly speaking, we must have
\begin{equation}\label{req}
 \frac{m_S}{\sqrt{|\lambda_{\phi \sigma}|}} \sim 10 GeV. 
\end{equation}

As we are taking only perturbative mixing coupling values, $\lambda_{\phi \sigma} < 4\pi$ and so $M_S \lesssim 35$~GeV. As we can infer from figure~\ref{sigmaMsig}, this possibility is quite excluded by direct detection bounds.

\begin{figure}[!htb]
\minipage{0.35\textwidth}
  \includegraphics[width=\linewidth]{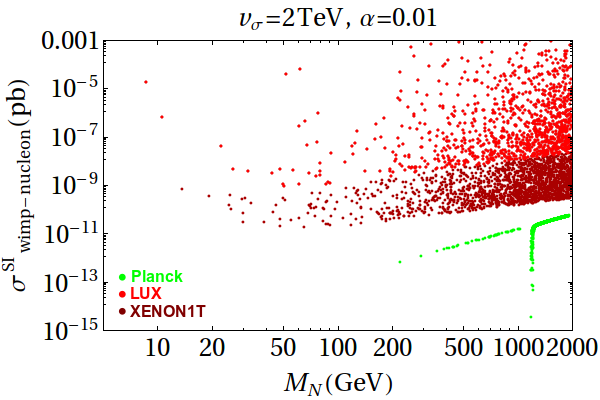}
\endminipage 
\minipage{0.35\textwidth}
  \includegraphics[width=\linewidth]{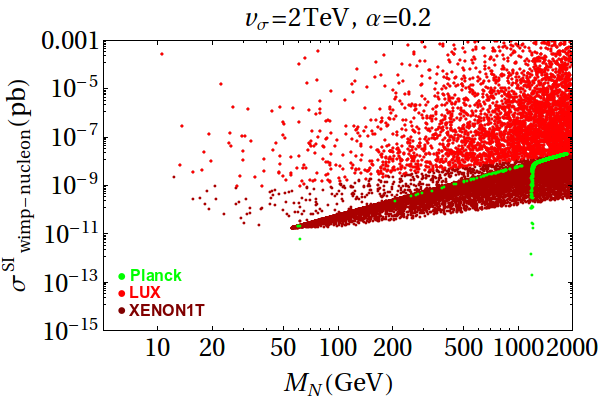}
\endminipage
\minipage{0.35\textwidth}
  \includegraphics[width=\linewidth]{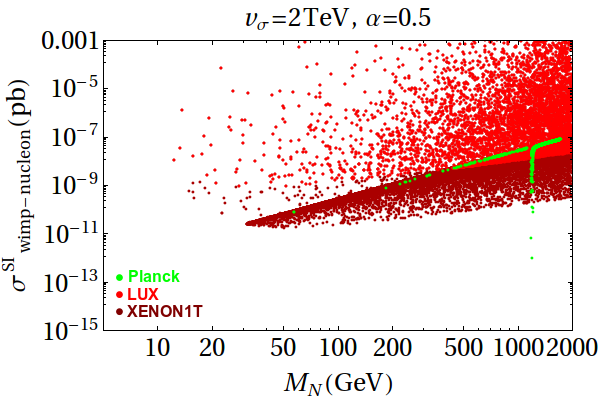}
\endminipage
\caption{Scattering cross section as a function of the Majorana fermion for certain values of mixing angle. The Planck-allowed points are for a heavier than Higgs singlet scalar.}
\label{anglecross}
\end{figure}

Finally, in figure~\ref{anglecross} we see that we can evade the direct detection bounds for a small enough mixing angle (left), as the interacion of the Majorana fermion with the standard particles is drove by a Higgs portal. This was also observed in ref.~\cite{Fairbairn:2013uta}. Again, in the scalar degenerate case, when $M_S \sim 125$~GeV, we can also evade the direct detection bounds, but the XENON1T future results should constrain strongly this case as well. In these plots, such a scenario has Planck-allowed points at $M_N \approx 1.2$~TeV. All this analysis is a fair update concerning the issue of DM relic density and direct detection. Once we have a lot of information since the Higgs boson discovery we can use it to further improve our results on the parameter space siege for this model, the so called collider complementarity. We perform this task in the following section. 

In summary, we have restricted our parameter space by considering only coupling constants within the perturbative regime and ensuring a potential bounded from below. We have considered mixing angles compatible with the maximum mixing allowed for the discovered Higgs boson and examined the impact of the future measurements of the Higgs self-couplings on our model. The values for $v_\sigma$ were limited from thousands of GeV to few TeV in order to furnish the observed relic density and we have seen that our model is not capable of explaining the existence of a very light singlet scalar if we consider it related to the dark matter physics, through the direct detection bounds. Now, we are ready to look for viable regions of our parameter space taking the experimental bounds in a complementary way.

\section{Complementarity bounds}\label{s4}

This model predict new decay channels for the Higgs boson, that are so far invisible for the LHC. The new decay rates are given by:

\begin{itemize}
 \item $\Gamma(H \rightarrow JJ) = \frac{M_H^3 s_\alpha^2}{32 \pi v_\sigma^2}$
 \item $\Gamma(H \rightarrow NN) = \frac{1}{16 \pi} \frac{s_\alpha^2 M_N^2 M_H}{v_\sigma^2} \left(1-\frac{4M_N^2}{M_H^2}\right)^{3/2}$
 \item $\Gamma(H \rightarrow SS) = \frac{M_H^3}{32 \pi}\frac{s_\alpha^2 c_\alpha^2}{v_\sigma^2} \left(1 + \frac{2 M_S^2}{M_H^2}\right)^2\left(c_\alpha + s_\alpha\frac{v_\sigma}{v_\phi}\right)^2\sqrt{1-\frac{4M_S^2}{M_H^2}}$
\end{itemize}

\vspace{.2cm}
Considering the Higgs decay rate in SM as $\Gamma^{SM} \approx 4$~MeV~\cite{Khachatryan:2014jba}, the upper bound on the branching ratio for the invisible Higgs decay (IHD)~\cite{Zhou:2014dba} is 
\begin{equation}
 B_H^{inv} = \frac{\Gamma^{inv}}{c_\alpha^2 \Gamma^{SM} + \Gamma^{inv}} < 0.4,
\end{equation}
and give us the following constraint on our free parameters:

\begin{equation}
 \frac{\Gamma^{inv}}{c_\alpha^2} < 2.67 \times 10^{-3} GeV.
\end{equation}
  
As we focus in the heavier than Higgs singlet scalar case, there are only two possible hierarchies: $M_H < 2 M_S, 2 M_N$ and $2M_N < M_H < 2M_S$. For $\Gamma^{inv} = \Gamma(H \rightarrow JJ)$, we have 
\begin{equation} \label{hjj}
 \frac{tg^2_\alpha}{v_\sigma^2} < 1.4 \times 10^{-7} 
\end{equation}
and for $\Gamma^{inv} = \Gamma(H \rightarrow JJ) + \Gamma(H \rightarrow NN)$,
\begin{equation} \label{hjjnn}
 \frac{tg^2_\alpha}{v_\sigma^2} \left[ 1 + 2 \frac{M_N^2}{M_H^2} \left(1-\frac{4M_N^2}{M_H^2} \right)^{3/2}  \right] < 1.4 \times 10^{-7}.
\end{equation}

As the Z boson do not decay at tree level in any singlet scalar in this model, the constraints from invisible Z decay is not relevant here. 

In order to explore the regions of our parameter space allowed by the Planck, LUX, XENON1T and IHD constraints, we will choose values of the free parameters based on the study of the previous section, instead of arbitrary ones. We do this next by fixing first pairs of free parameters that parametrize our scalar sector extension, $v_\sigma$ and $\alpha$, and then pairs of mass values of the new particles of our spectrum, $M_N$ and $M_S$. 

For all the following free parameter choices, we have checked that the coupling constants $\lambda_N$ is always smaller than unity. Also, all the viable parameter space regions are consistent with $\lambda_{\phi \sigma} < 1$ because of the direct detection bounds.

In what follows, we will identify the regions excluded by the invisible Higgs decay bounds. The regions filled with blue straight lines are in the hierarchy $M_H < 2M_N, 2M_S$ and do not obey the eq.~\ref{hjj}. The regions filled with orange straight lines are in the hierarchy $2M_N < M_H < 2M_S$ and do not obey the eq.~\ref{hjjnn}. Consistently with our previous figures, the green lines are composed of points within the Planck interval giving a sufficiently abundant Majorana fermion, while the gray regions are composed by the points that give underabundant Majorana fermion, indicating the regions of the parameter space in which the Majorana fermion could be only a fraction of the total DM content of the Universe. Although we do not consider another DM candidate or non-thermal production that validate these gray points, we keep it for completeness. Regarding the direct detection search, the red regions are excluded by LUX and the dashed dark red lines will extend this exclusion if XENON1T do not confirm any 
DM signal in a close future.

\subsection{$v_\sigma$ and $\alpha$ fixed}

In this subsection, we fix a set of values for the new vev, $v_\sigma$, and the mixing angle, $\alpha$, that parametrize the scalar extension of this model. Based on our previous results, lustrated by figure~\ref{vev}, we will study how the lepton number breaking scale changes the allowed regions by choosing a vev $v_\sigma =$ 500~GeV, 1~TeV and 2~TeV. The role of the mixing angle will be considered by taking it as $\alpha =$ 0.01, 0.2 and 0.5. 

\begin{figure}[!h]
\minipage{0.5\textwidth}
  \includegraphics[width=\linewidth]{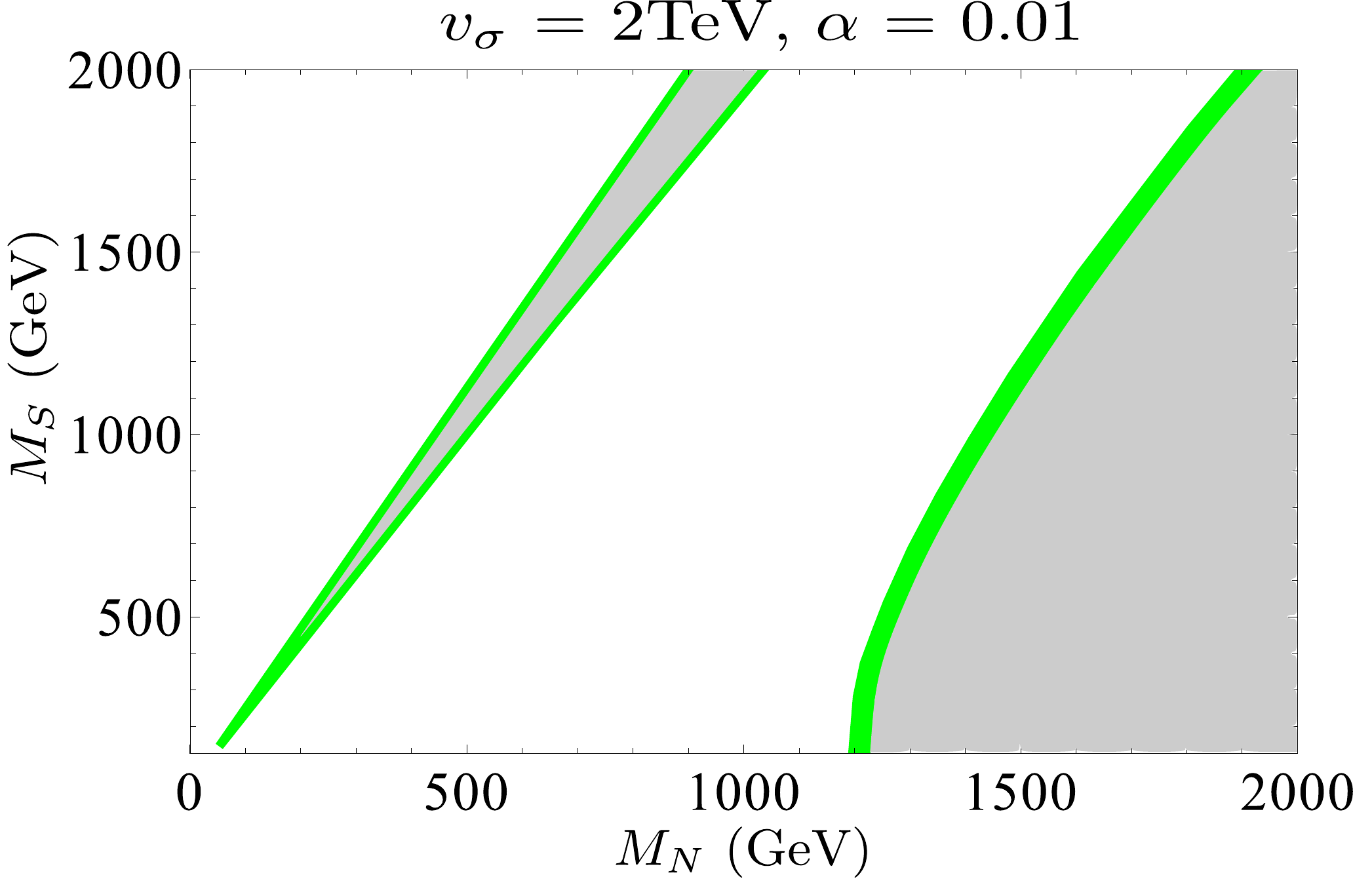}
  \endminipage\hspace{0.5cm}
\minipage{0.5\textwidth}
  \includegraphics[width=\linewidth]{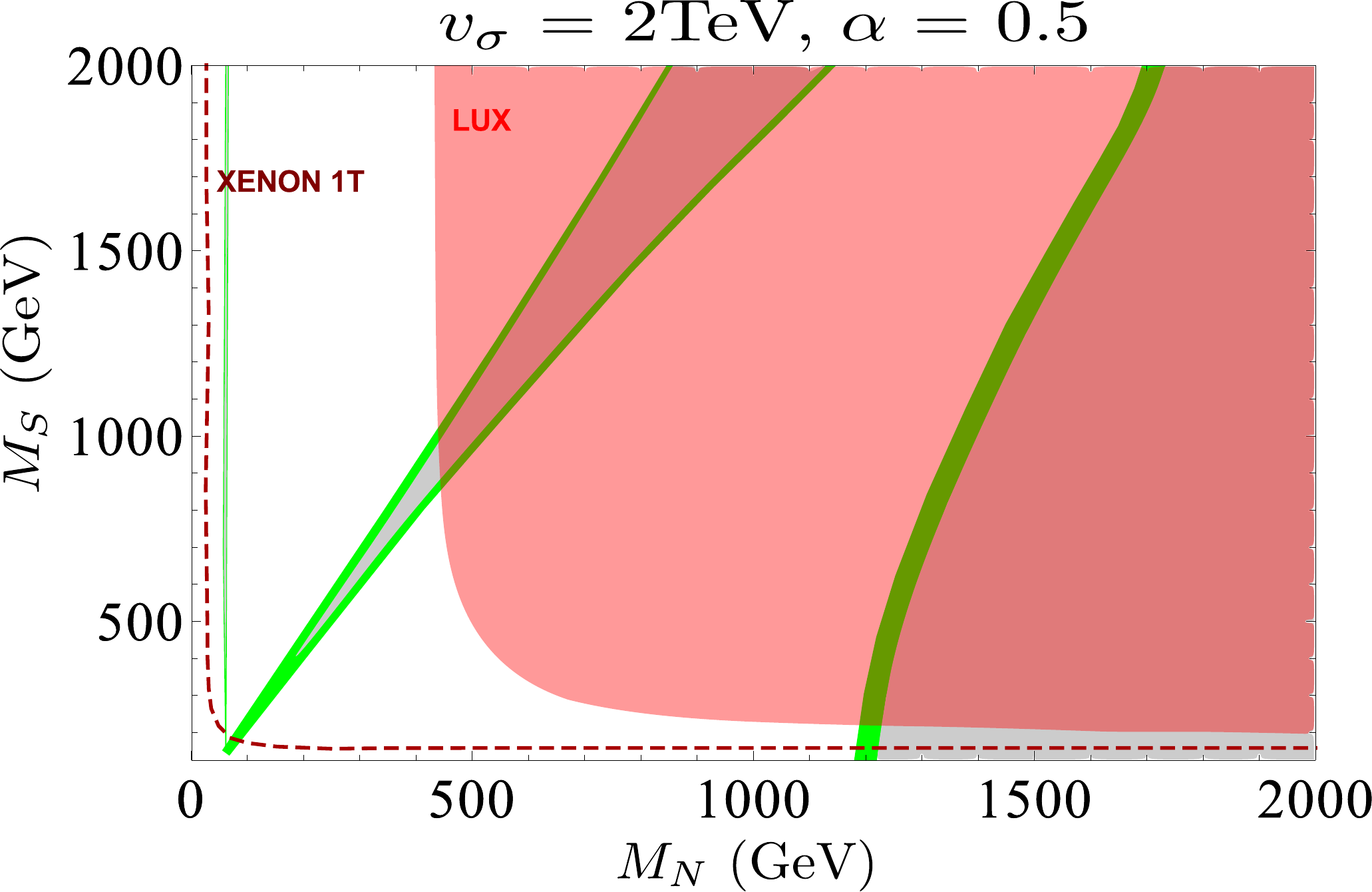}
  \endminipage
\begin{center}
\minipage{0.5\textwidth}
  \includegraphics[width=\linewidth]{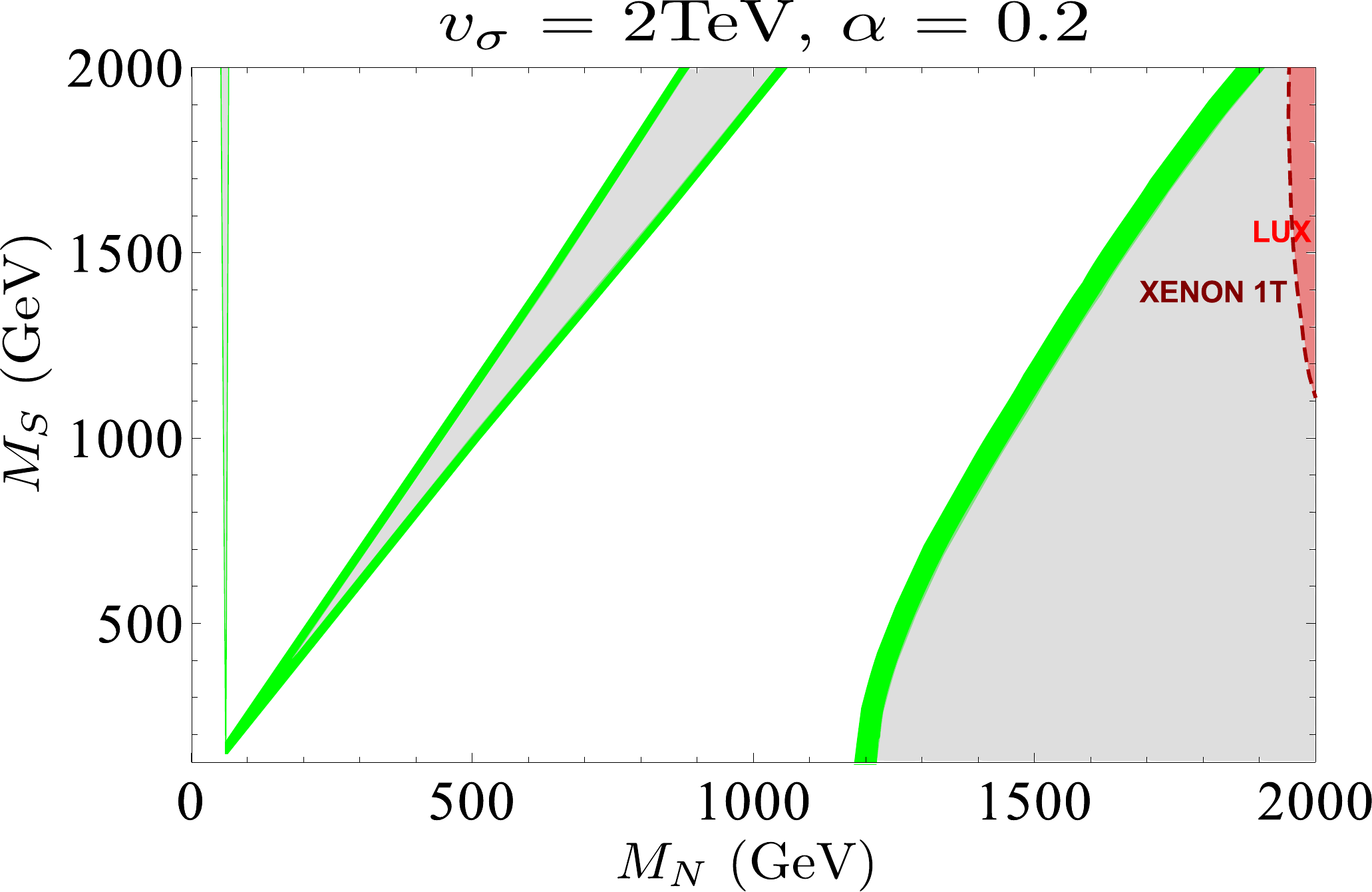}
  \endminipage
\end{center}
\minipage{0.5\textwidth}
  \includegraphics[width=\linewidth]{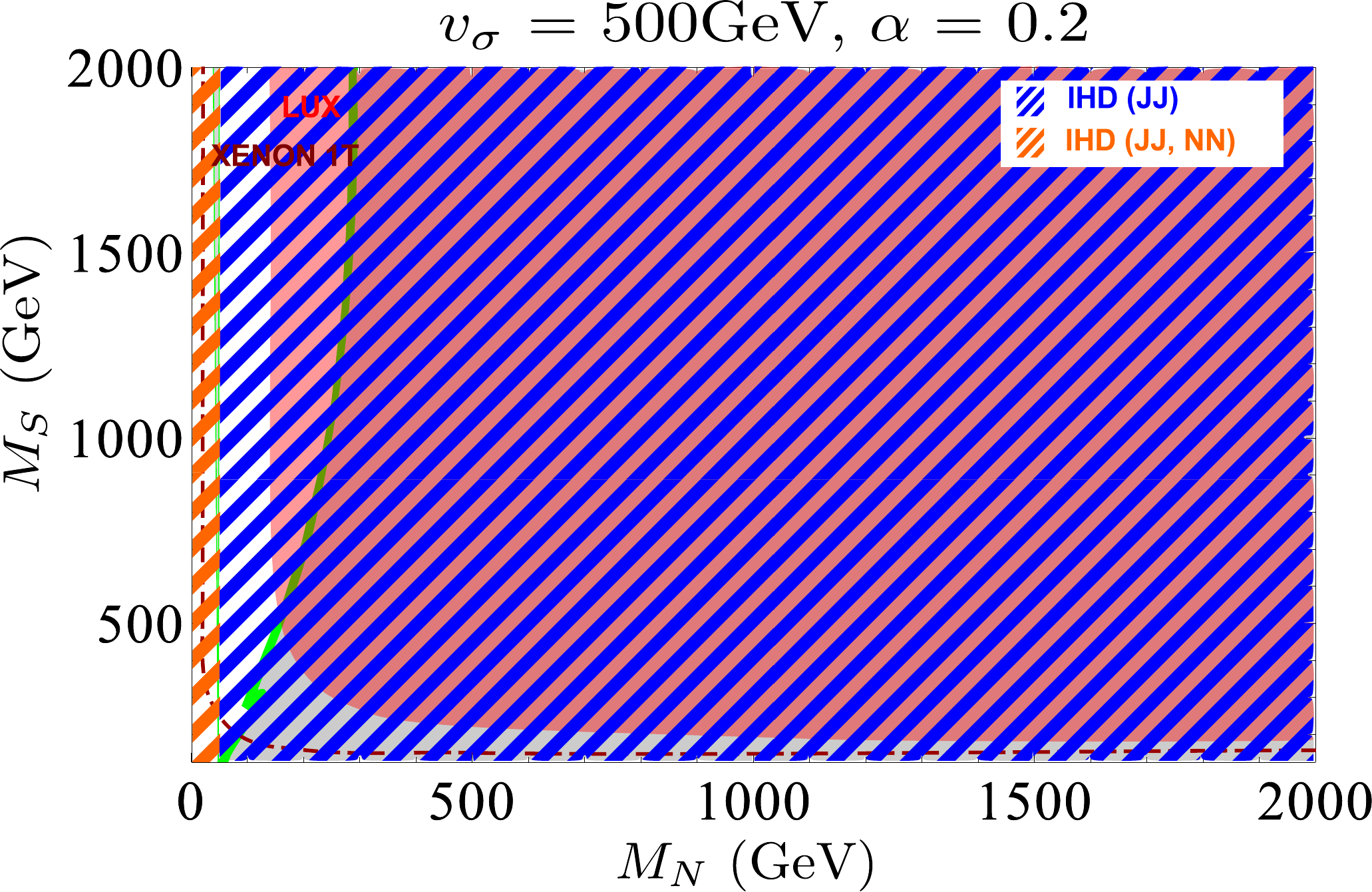}
  \endminipage\hspace{0.5cm}
\minipage{0.5\textwidth}
  \includegraphics[width=\linewidth]{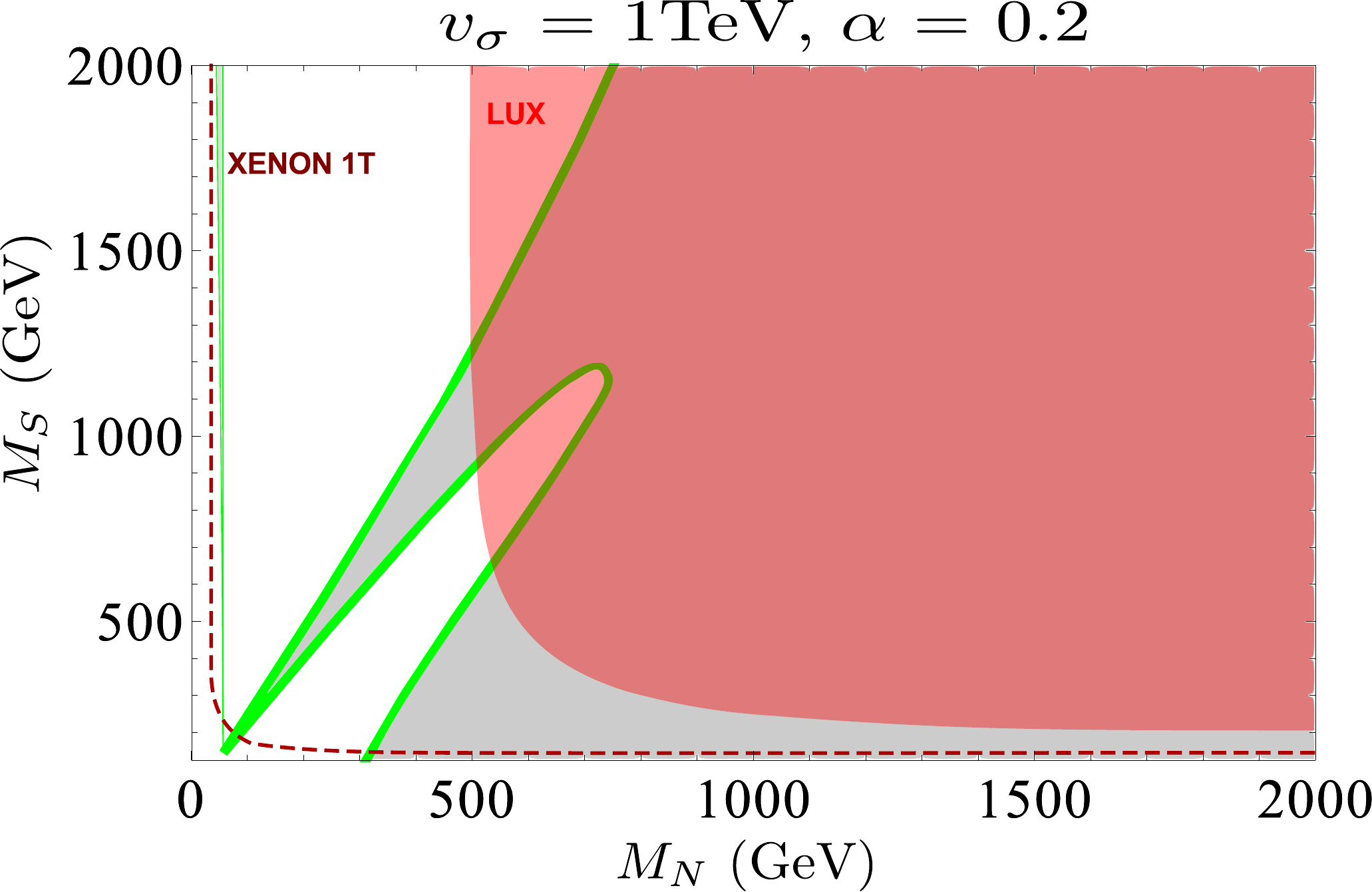}
  \endminipage
\caption{Free parameter slices ($M_N, M_S$) for fixed $v_\sigma$ values (upper panel) and fixed $\alpha$ values (lower panel). The green region are in agreement with Planck while the gray one give underabundant Majorana fermion. The red region are already excluded by LUX and the dark red dashed lines shows how XENON1T should expand this exclusion if do not observe a DM signal. The blue (orange) straight lines show the region excluded by IHD into two majorons (two majorons and two Majorana fermions).}
\label{mnmsig}
\end{figure}

In figure~\ref{mnmsig}, we display the free parameter slice $(M_N, M_S)$ for different pairs of $\alpha$ and $v_\sigma$. All the Planck-allowed points (green) are near the scalar resonance, in agreement with previous works. For strong enough scalar mixing ($\alpha=0.2$ and $\alpha=0.5$ in this figure) we see Planck-allowed points for $M_N \sim M_H/2 = 62.5$~GeV, corresponding to the Higgs resonance. As we are varying $M_S$ up to 2~TeV, in the three first plots (for $v_\sigma = 2$~TeV) the Planck-allowed points for $M_N < 1$~TeV correspond to the singlet scalar resonances and the ones for $M_N > 1$~TeV, to the falls of the abundance curve (see figure~\ref{resonances}). 

For $\alpha = 0.01$ and $\alpha = 0.2$, the direct detection search do not exclude any Planck-allowed point, but for $\alpha = 0.5$ it offers a strong limit. We see that $M_N>500$~GeV is excluded by LUX for $\alpha = 0.5$ if $M_S>500$~GeV and if XENON1T do not confirm any signal, this case becomes practically excluded. For a TeV scale vev, all the points are allowed by the IHD, but $v_\sigma=500$~GeV (left lower panel) is completely excluded by IHD constraint, considering a conservative mixing angle of 0.2. For a $v_\sigma = 1$~TeV (right lower panel), we see that $M_N>500$~GeV is also excluded by LUX for $\alpha = 0.5$ if $M_S>500$~GeV but in comparison with the case ``$v_\sigma = 1$~TeV, $\alpha = 0.5$'' (right upper plot), we have more Planck-allowed points in the LUX-allowed region.

In the limit of null mixture ($\alpha \rightarrow 0$) and very high energy breaking scale of the symmetry we are supposing ($v_\sigma \rightarrow \infty$), the scalar sector of our model becomes standard, with just one scalar coming from a doublet, no Majoron in the spectrum and without the possibility of provides mass for our dark matter candidate. As we have seen in figure~\ref{mnmsig}, our parameter space is favored by small deviations of the scalar sector, it means, for smaller scalar mixing and very high energy physics (left upper panel), in view of the current collider and direct detection searches. We have seen that the model we consider here is very predictive and in the next few decades, all this scenario will be tested by the experiments. It is constructive to keep in mind, however, that the confirmations of a possible DM particle signal \textit{and} a possible new scalar discovery at colliders or accelerators are very connected things for Higgs portal models and their immersion in more complex 
contexts may offer important hints.

\subsection{$M_N$ and $M_S$ fixed}

Now we study the allowed parameter space for fixed values of $M_N$ and $M_S$. Here we consider the lepton number breaking scale, $v_\sigma$, in a larger window, from 100~GeV to 10~TeV. The TeV scale is a natural scale in the search for new physics, so we study separately the case of $M_N=1$~TeV and $M_S=1$~TeV. For a $M_N=1$~TeV, we show how the allowed region depends on $M_S$ by fixing it at 130~GeV (quite degenerate case), 500~GeV (intermadiate case) and 1~TeV (TeV scale physics). On the other hand, for $M_S=1$~TeV, we consider Majorana fermions with masses of 60~GeV (Higgs resonance case) and 500~GeV (singlet scalar resonance case).
  
\begin{figure}[h]
\minipage{0.5\textwidth}
  \includegraphics[width=\linewidth]{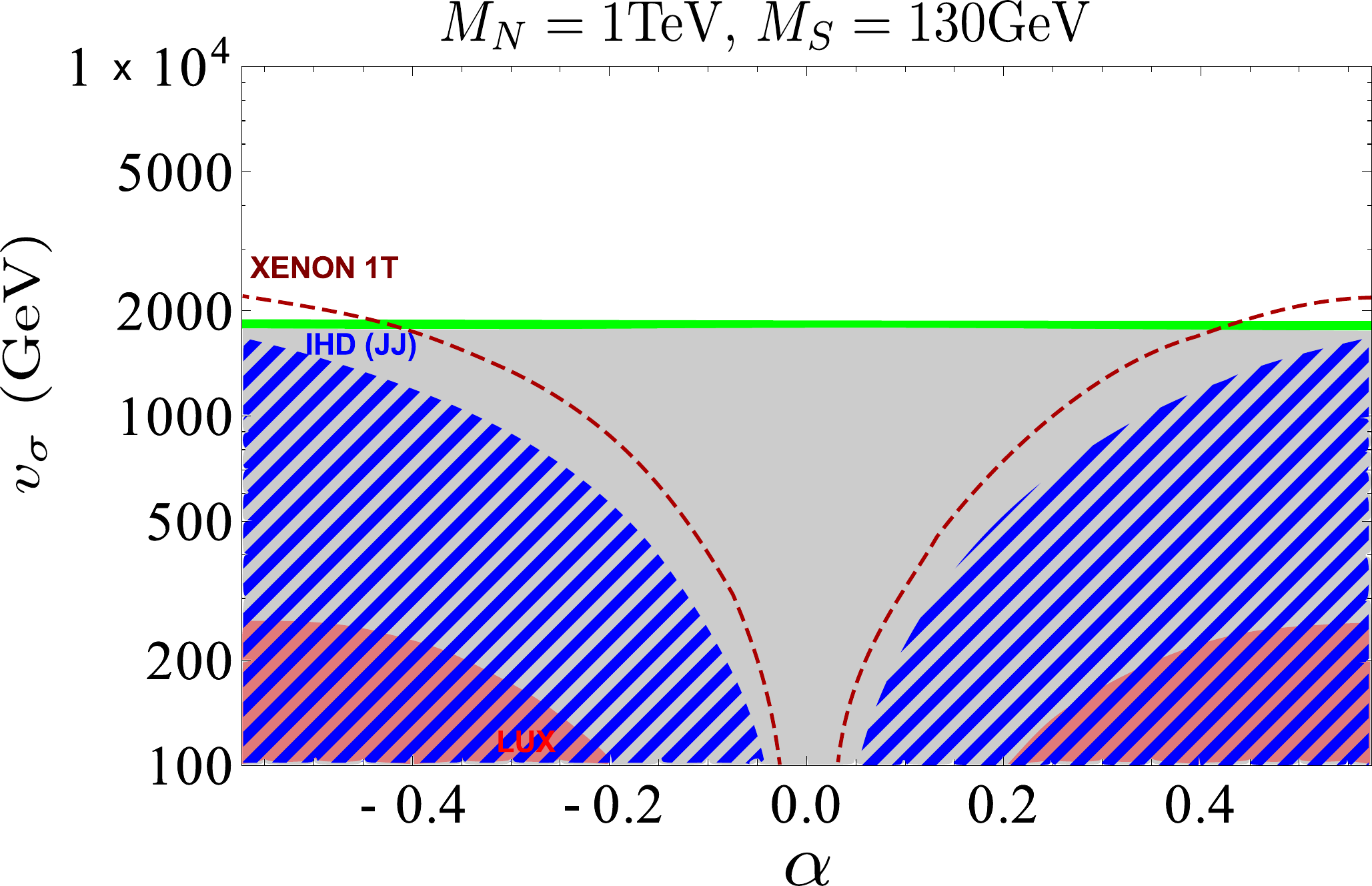}
  \endminipage \hspace{0.5cm}
\minipage{0.5\textwidth}
  \includegraphics[width=\linewidth]{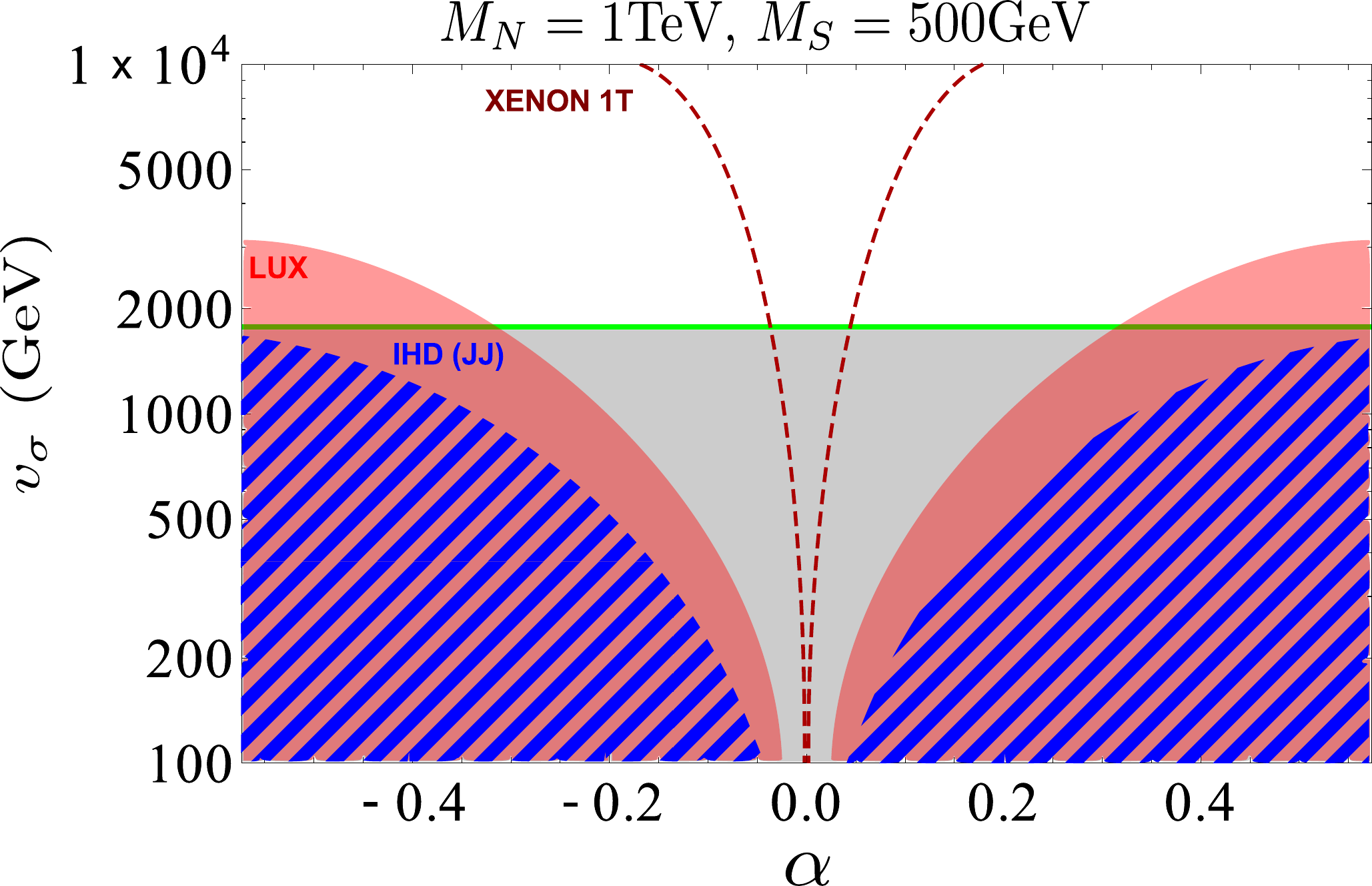}
  \endminipage
\begin{center}
\minipage{0.5\textwidth}
  \includegraphics[width=\linewidth]{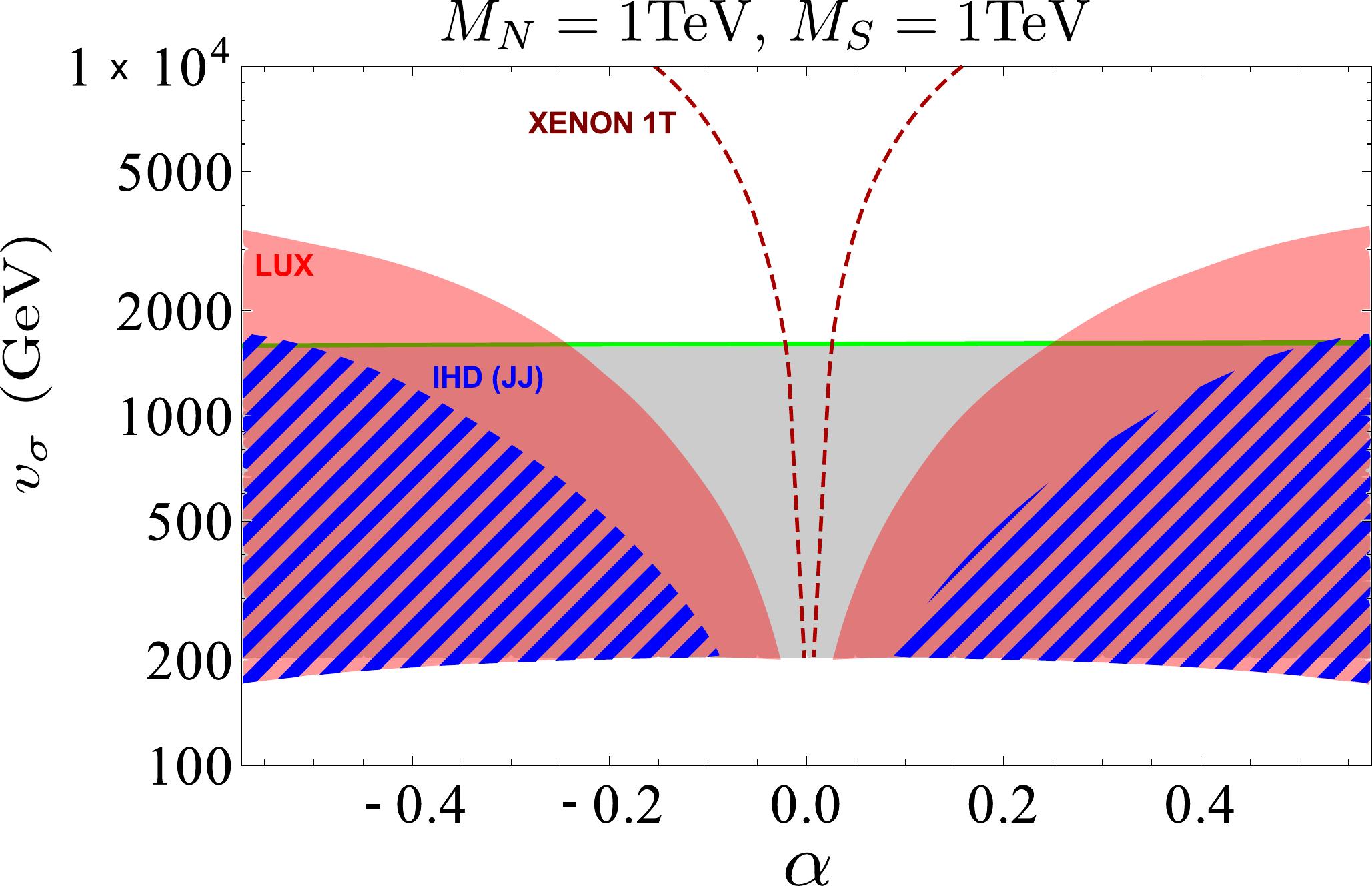}
  \endminipage
\end{center}
\minipage{0.5\textwidth}
  \includegraphics[width=\linewidth]{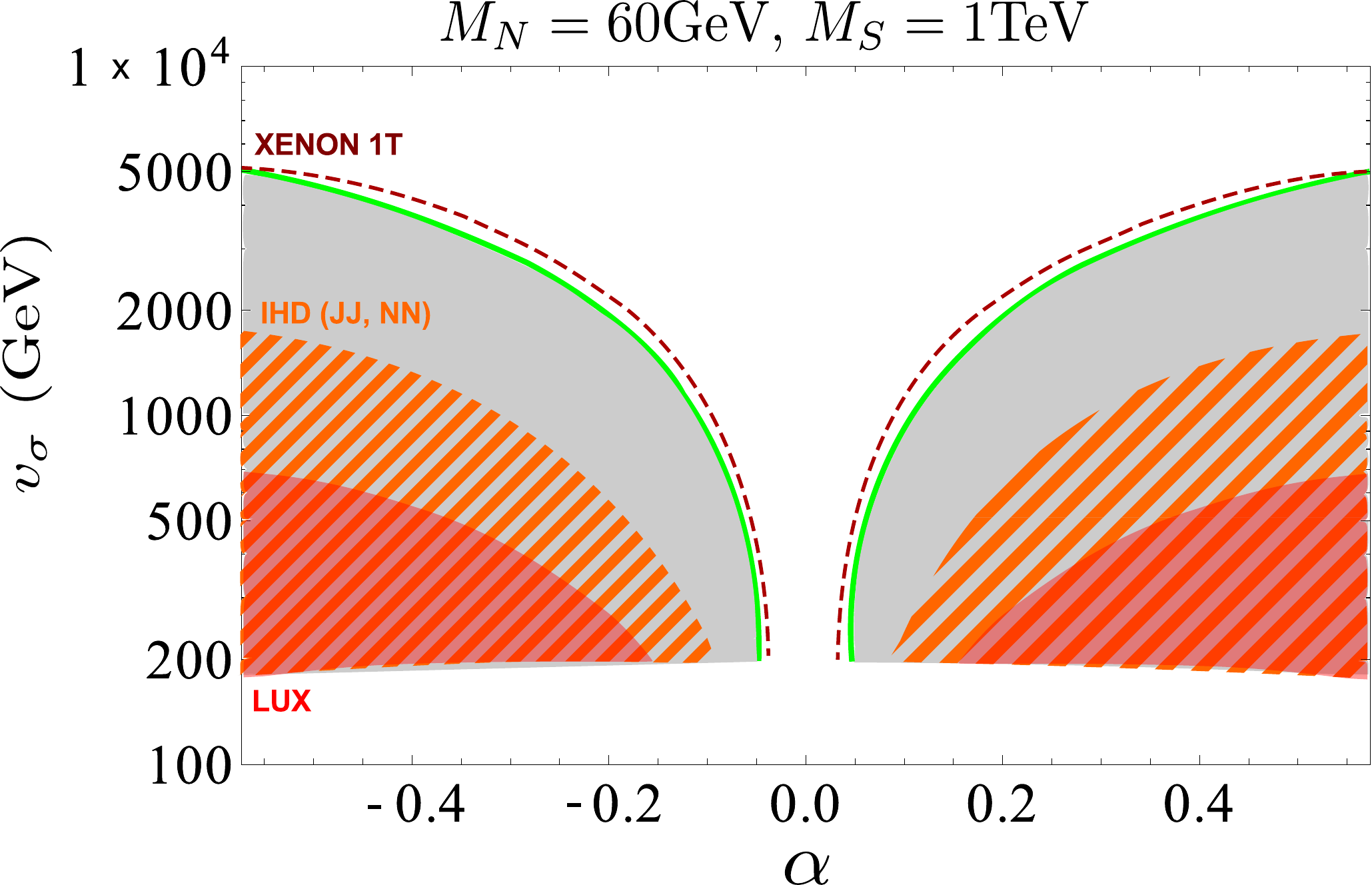}
  \endminipage\hspace{0.5cm}
\minipage{0.5\textwidth}
  \includegraphics[width=\linewidth]{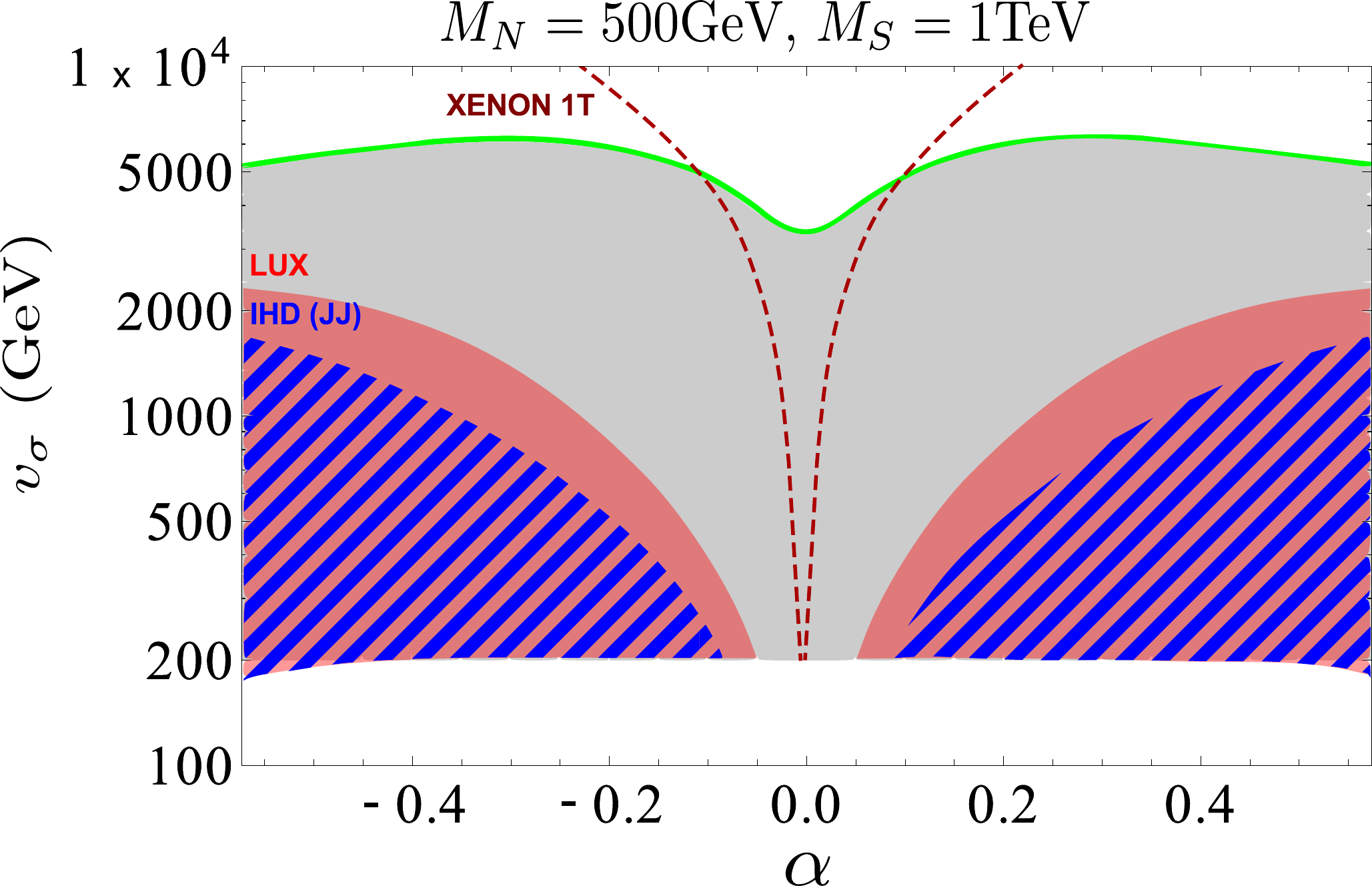}
  \endminipage
\caption{Free parameter slices ($\alpha, v_\sigma$) for fixed vev values (upper panel) and fixed mixing angle values (lower panel). Same color code of figure~\ref{mnmsig}.}
\label{alfavs}
\end{figure}

In figure~\ref{alfavs}, we show the free parameter slices ($\alpha,v_\sigma$). In the upper and middle panels, we have the case of $M_N=1$~TeV. The Majorana fermion will be sufficiently abundant only if the lepton number symmetry breaking is near 2~TeV, regardless the singlet scalar mass, implying that its interaction with the scalars would be weak (see eq.~\ref{mn}). Lower breaking scales would need another DM candidate or a non-thermal DM production, to be constrained by IHD, LUX and XENON1T. As we already pointed out, the quite degenerate case (left upper panel) can evade direct detection and as we may expect, the LHC provides the strongest current bound, but cannot constrain the scalar mixing of the Higgs with a scalar that develops vev too early, at TeV scale. The XENON1T will allow scalar mixings up to 0.4 in that case. For heavier singlet scalars, the current limits for the mixing are $\alpha \lesssim 0.3$ for $M_S=500$~GeV (right upper panel) and $\alpha \lesssim 0.2$ for $M_S=1$~TeV (middle panel). 
The XENON1T future limits will restrict much more the mixture, up to $\alpha \lesssim 0.05$, in which case non-standard interactions will not be detected easily.  

In the lower panel, we can appreciate how the scalar resonance regions of the spectrum evade the current experimental bounds. At left, we show the Higgs resonance case ($M_N = 60$~GeV), where the LHC bound is stronger than LUX bound but cannot constraint a sufficiently abundant Majorana fermion. In this case, the breaking scale should be from 200~GeV (for small mixings) to 5~TeV (for maximal mixing). Nevertheless, if no signal is confirmed by XENON1T, this case will be completely excluded in the next few years. At right, we show the singlet scalar resonance ($M_N = 500$~GeV), also not constrained by the current bounds but accessible to XENON1T, when the mixing will be restricted to $\alpha \lesssim0.1$. This possibility requires the breaking scale near 5~TeV for a sufficiently abundant Majorana fermion. 

We have seen that even the safer resonance regions will be strongly constrained in the near future, and even considering a TeV scale new physics.

\section{Conclusions}\label{s5}

In this work, we updated the Majorana fermion DM in a CP-conserving Higgs portal scenario, in view of the LUX, XENON1T and the invisible Higgs decay (IHD) recent results. 

Our dark matter candidate is a sterile neutral Majorana fermion and the Higgs portal is opened by a complex scalar field, singlet and neutral under the standard symmetries and charges. This singlet scalar develops spontaneously a vacuum expectation value (vev) that gives a Majorana mass for the fermion and leaves it stable due to the discrete $Z_2$ symmetry we are supposing, making it a DM candidate. This model is described by only four parameters: the Majorana fermion and singlet scalar masses, $M_N$ and $M_S$, the mixing angle between the scalars, $\alpha$, and the lepton number breaking scale, $v_\sigma$. 

Our parameter space is within the sensibility region of the future collider searches in what concerns to the 3-Higgs self-coupling constant. It means that the measurement of the Higgs self-coupling can shade light on our Majorana fermion DM scenario. Also, strong deviations from the SM value of the 4-Higgs self-coupling will give us information about the scalar mass hierarchy. All the analysis we have made is within the perturbative regime and is compatible with a bounded from below scalar potential. 

We found that $v_\sigma$ must be between few hundreds of GeV and few TeV in order to the Majorana fermion account for the total amount of DM in the Universe, according to the Planck results. Considering that this symmetry breaking took place at some scale between 100-2000GeV, the singlet scalar mass from 0.1-2000GeV and a mixing angle ensuring $|\cos(\alpha)|>0.84$, we can have a Majorana fermion as a DM candidate for masses from 30GeV to 2TeV. The case of a lighter than Higgs singlet scalar in this DM scenario is quite ruled out by LUX, making inviable to consider the Majoron as dark radiation. 

With the aim of ``split the degeneracy'' of this allowed region, we studied the role of each parameter to find the very predictive curves allowed by Planck and contrasted by the experimental limits. We did it systematically by choosing benchmark values for the free parameters in two steps. Firstly we examined the scalar extension parametrization, by fixing $\alpha$ at 0.01, 0.2 and 0.5 and $v_\sigma$ at 500GeV, 1TeV and 2TeV. Secondly, we examined the mass spectrum, by fixing $M_N$ at 60GeV, 500GeV and 1TeV and $M_S$ at 130GeV, 500GeV and 1TeV. 

As expected, the allowed regions are near the scalar resonances. We have shown that it is possible to evade the LUX and even the XENON1T constraint for a Majorana DM candidate in two situations: scalar degenerate case (fig.~\ref{msigfixed}) and small mixing angles (fig.~\ref{anglecross}). 

The case of a relatively low breaking scale ($v_\sigma=$~500GeV) was found totally excluded by IHD even for a mixing angle of only 0.2 and TeV scale $v_\sigma$ values were found safe (fig.~\ref{mnmsig}). We have shown that $M_N > 500$~GeV is excluded by LUX if $M_S > 500$~GeV in essentially two situations: ``$v_\sigma = 2$~TeV,$\alpha=0.5$'' and ``$v_\sigma=1$~TeV,$\alpha=0.2$'' (right upper and lower of fig.~\ref{mnmsig}, respectively).

For $M_N=1$~TeV, $v_\sigma$ must be at approximately 2~TeV in order to agree with the Planck results and the mixing was found restricted to $\alpha \lesssim 0.3$ (for $M_S=500$~GeV) and $\alpha \lesssim 0.2$ (for $M_S=1$~TeV) by the LUX results and to $\alpha \lesssim 0.05$ by the XENON1T future bound (upper and middle panels of fig.~\ref{alfavs}). It means that if no signal is confirmed by XENON1T, it becomes very difficult to detect some non-standard interaction at colliders, since they should be proportional to $\sin(\alpha)^2$. The resonance regions were found not constrained by the current experiments, but XENON1T will exclude completely $M_N \sim 60$~GeV and restrict the mixing to $\alpha \lesssim 0.1$ for $M_S=1$~TeV (lower panel of fig.~\ref{alfavs}).

We have seen how the discoveries of a DM and a new scalar particles are strongly related in our Higgs portal scenario. While this minimal scheme remains allowed by the experimental bounds, it can be embedded in more fundamental extensions of the SM. In supersymmetric theories, stable neutralinos are Majorana fermions that also appear as natural dark matter candidates and in some gauge extensions and neutral scalars may be the remnants of a more fundamental gauge symmetry breaking, as studied in~\cite{Dong:2014wsa}.\\

\noindent {\bf Acknowledgments:}\\
We are very grateful to Clarissa Siqueira, Antonio Oliveira and Farinaldo Queiroz for useful discussions. This work was supported by Coordenação de Aperfeiçoamento de Pessoal de Nível Superior - CAPES (MD) and Conselho Nacional de Desenvolvimento Científico e Tecnológico - CNPq (CASP,PSRS).\\

\noindent{\bf References} \\ 
\providecommand{\href}[2]{#2}\begingroup\raggedright

\endgroup

\end{document}